\documentclass[a4paper,12pt]{article}
\pdfoutput=1 

\usepackage{jheppub} 

\usepackage[T1]{fontenc} 
\usepackage{amsmath,amssymb}
\usepackage{dsfont}
\usepackage{upgreek}
\usepackage{slashed}
\usepackage{appendix}
\newcommand{\be}{\begin{equation}}
\newcommand{\ee}{\end{equation}}
\newcommand{\ben}{\begin{eqnarray}}
\newcommand{\een}{\end{eqnarray}}
\newcommand{\nn}{\nonumber}
\newcommand{\p}{\partial}
\def\l{\left}
\def\r{\right}

\title{\bf{On KKLT/CFT and LVS/CFT Dualities}}
 \author[a]{\small{Senarath de Alwis},}
\author[b]{\small{Rajesh Kumar Gupta},}
\author[b,c]{\small{Fernando Quevedo},}
\author[b,d]{\small{Roberto Valandro}}
\affiliation[a]{\small{UCB 390, Physics Department, University of Colorado, Boulder CO 80309, USA}}
\affiliation[b]{\small{ICTP, Strada Costiera 11, 34151 Trieste, Italy}}
\affiliation[c]{\small{DAMTP, CMS, University of Cambridge, Wilberforce Road, Cambridge, CB3 0WA, UK.}}
\affiliation[d]{\small{Dipartimento di Fisica dell'Universit\`a di Trieste and INFN, Sezione di Trieste,
Strada Costiera 11, 34151 Trieste, Italy.
}}
\emailAdd{dealwiss@colorado.edu}
\emailAdd{rgupta@ictp.it}
\emailAdd{f.quevedo@damtp.cam.ac.uk}
\emailAdd{rvalandr@ictp.it}

\abstract{\small{We present a general discussion of the properties of three dimensional CFT duals to the AdS string theory vacua coming from type IIB Calabi-Yau  flux compactifications. Both KKLT  and Large Volume Scenario (LVS) minima are considered. In both cases we identify the large `central charge', find a separation of scales between the radius of AdS and the size of the extra dimensions and show that the dual CFT has only a  limited number of operators with small conformal dimension. Differences between the two sets of duals are identified. Besides a different amount of supersymmetry ($\mathcal{N}=1$ for KKLT and $\mathcal{N}=0$ for LVS) we find that the LVS CFT dual has only one scalar operator with $\mathcal{O}(1)$ conformal dimension, corresponding to the volume modulus,  whereas in KKLT  the whole set of  $h^{1,1}$ K\"ahler moduli have this property.  Also, the maximal number of degrees of freedom is estimated to be larger in LVS than in  KKLT duals. In both cases we explicitly compute the coefficient of the logarithmic contribution to the one-loop vacuum energy which should be invariant under duality and therefore provides  a non-trivial prediction for the dual CFT. This coefficient takes a particularly simple form in the KKLT case. }}

\preprint{DAMTP-2014-92} 

\begin{document} 
\maketitle
\flushbottom

\section{Introduction}

Flux compactifications of type IIB string theory have given rise to two major developments within string theory: AdS/CFT duality \cite{Maldacena:1997re,Witten:1998qj} (see \cite{adscft} for a review) and the string landscape \cite{Grana:2005jc,landscape,Blumenhagen:2006ci,Dasgupta:1999ss,gkp,kklt,Douglas:2003um,Ashok:2003gk,Denef:2004dm,DenefDouglas1,DenefDouglas2,Acharya:2005ez} of moduli stabilised four dimensional (4D) string vacua. In the simplest cases,
these four dimensional minima have a negative cosmological constant and hence are 
AdS$_4$ vacua. It is then
natural to inquire if these Anti de Sitter (AdS) vacua of the string landscape have Conformal Field Theory (CFT)
duals and if so what the properties of these theories are.


Identifying CFT duals of the AdS (and dS) vacua of the string landscape
would be a way to provide a proper non perturbative description of these
vacua and put the string landscape on firmer ground. This is the subject
of the present article. For previous discussions of this issue see
\cite{eva, evajoe,banks,vafa, aharony, denef,torroba}.\footnote{AdS$_{d+1}$/CFT$_{d}$ duality has also been used in Calabi-Yau flux compactifications in a different context that should not be confused with our target in this article. In those cases, conifold geometries such as the Klebanov-Strassler warped throat are embedded in compact Calabi-Yau manifolds and provide a stringy realisation of the Randall-Sundrum set-up with the tip of the throat providing the IR brane and the compact Calabi-Yau at the beginning of the throat providing the UV Planck brane \cite{Verlinde:1999fy}. In these cases AdS$_{d+1}$/CFT$_{d}$ duality is used in the sense that 4D field theories are dual to 5D gravity theories in which locally the five dimensions are the 4D spacetime dimensions plus the direction along the throat, i.e. $d=4$. On the other hand, in this paper we are concentrating on three-dimensional field theories dual to four-dimensional gravity theories, i.e. $d=3$.}

By now there are two main scenarios of moduli stabilisation in type IIB string compactifications on Calabi-Yau (CY) manifolds: KKLT  \cite{kklt}\, and the Large Volume Scenario  (LVS)~\cite{lvs}. 
Contrary to the original  AdS$_5\times S^5$ background where the flux was enough to stabilise the geometric modulus of $S^5$, in KKLT and LVS scenarios the fluxes fix only part of the geometric moduli (this can be read from the ten dimensional equation of motions \cite{Dasgupta:1999ss,gkp}, like for AdS$_5\times S^5$) leaving some flat directions. A key ingredient to stabilise the remaining geometric moduli (in a  AdS$_4$ vacuum) is the presence of  non-perturbative effects in the 4D effective field theory (EFT) obtained after compactification. This makes a full ten dimensional (10D) analysis of these vacua very difficult and we can only rely on the EFT results. Black-brane solutions that were at the origin of the AdS$_5\times S^5$/CFT$_4$ duality are not available and therefore there is less control on the potential duality in the KKLT and LVS cases. This explains the relative shortage of efforts to study  the CFT duals of these vacua  during the past ten years.
Another difference with  AdS$_5 \times S^5$ is that  in both KKLT and LVS scenarios there is a hierarchy between the size of the internal dimensions and the
AdS radius. This is in contrast to the situation in Freund-Rubin compactifications where one needs to establish on a case by case that there is a consistent truncation to the massless modes of the KK tower (see for example the discussion in section 2.2..5 of \cite{adscft}) . 

Even though both KKLT and LVS are based on Calabi-Yau flux compactifications of type IIB string
theory down to 4D, they have important differences that should be reflected in the dual CFTs.

\begin{itemize}

\item
The two scenarios realise the
separation of scales that allow the neglect of part of the spectrum in different ways. In KKLT this happens because of the small value of the flux
superpotential, while in LVS because of the hierarchically large value of the volume of the compactification manifold. In fact, KKLT relies on the possibility of tuning the flux superpotential $W_{flux}$ to very small values (of the same order of the non-perturbative superpotential), while LVS  is based on a generically order one $W_{flux}$. 

\item
The KKLT AdS$_4$ vacuum preserves $\mathcal{N}=1$ supersymmetry, whereas the LVS AdS$_4$ vacuum breaks
 supersymmetry spontaneously, with the breaking  being induced by generic 
fluxes. 

The fact that the LVS vacuum is not supersymmetric may raise concerns regarding its
stability and the existence of a CFT dual. It was shown in \cite{stability} that
as long as the effective field theory is valid the corresponding vacua are
stable under bubble nucleation and therefore a dual CFT is expected to
exist. Moreover, the fact that supersymmetry is spontaneously broken on the AdS side raises
the question of how this breaking manifests itself on the CFT side.

\item
Both scenarios allow the possibility to extend the AdS compactifications
to include dS. However, they are usually realised in different ways in both scenarios.\footnote{See \cite{kklt,compact,Louis:2012nb,compact2,compact4} for explicit dS minima in the type IIB context considered in this paper.} 
Addressing the possibility of duals to these dS vacua is very relevant, but
since these vacua are more model dependent and the dS/CFT duality is less understood we will not
address this issue here. Our discussion here may be relevant for a future
approach to this question.

\end{itemize}


In this article we make a general discussion of this potential duality with the intention to learn as much as possible about the properties of the CFT$_3$ duals. We are aware of the difficulty of the task and attempt only to extract general properties of the CFT$_3$. Motivated by the recent works on the black holes and AdS/CFT \cite{Banerjee:2010qc,Banerjee:2011jp,Sen:2011ba,Gupta:2014hxa, Chowdhury:2014lza,Bhattacharyya:2012ye,Beccaria:2014qea}, we compute the one loop partition function in supergravity and extract the universal contribution to the free energy. The universal contribution is proportional to the logarithm of the size of the AdS space and will correspond to $\log c$-correction to the free energy of the dual CFT. We  carry out the computation of this universal quantity on the AdS side of both the KKLT and LVS compactifications. As we will explain later, to do these computations, we work in a limit in which we only keep the contributions from massless supergravity fields and  K$\ddot{\text a}$hler moduli and  ignore the contribution from the complex structure moduli and dilaton (which have been supersymmetrically stabilized at a high scale) and the KK fields. Also in this limit the computation of universal contribution reduces to the calculation of the coefficient of $\ln|W_0|^2$. The result of this coefficient for the case of KKLT and LVS case are given in (\ref{KKLTGammaW01}) and (\ref{LVSGammaW01}) respectively. Because of supersymmetry in the case of KKLT, the expression of this coefficient is much simpler and can be expressed in terms of conformal dimension of operators dual to massive K$\ddot{\text a}$hler moduli. Being universal, the result of this calculation should provide a consistency check for any candidate CFT dual.

We organise this paper as follows. In Section \ref{sec:AdSFromFlux} we will present a detailed comparison between AdS$_5\times S^5$ background and the Calabi-Yau flux compactifications. 
In Section \ref{sec:PropertiesCFT3} we describe some properties of the three dimensional CFT dual to KKLT and LVS flux compactification. In particular we identify the amount of supersymmetry, the central charge, the conformal dimension of the various operators dual to fields on the gravity side and the baryonic operator/vertex in the dual CFT. In Section \ref{sec:EffPot} we discuss the one loop corrections to the partition function in supergravity. These corrections will correspond to $\frac{1}{N}$ effects in the partition function of the dual CFT.  In this computation we calculate the above mentioned universal contribution to the partition function of the dual CFT and discuss the limit in which we perform the computation. In Section \ref{sec:logW0} we explicitly compute this term in the KKLT and LVS cases. This gives a prediction for the universal contribution to the partition function of the dual CFTs.

\section{AdS backgrounds from flux compactifications}\label{sec:AdSFromFlux}

The bosonic part of the 10D supergravity effective action for type IIB string theory in the Einstein frame is
\be
S=\frac{1}{(2\pi)^7\alpha'^4}\; \int d^{10}x\sqrt{-g}\left\{{\mathcal R}-\frac{\partial_M S\partial^M \bar{S}}{2(\rm{Re} S)^2}-\frac{G_3 \cdot {\bar G}_3}{12\rm{Re} S}-\frac{F_5^2}{4\cdot 5!}\right\}+ S_{CS}+S_{loc} \:.
\ee
Here $S=e^{-\phi}+iC_0$ is the axiodilaton field, $G_3=F_3-iSH_3$ the complex combination of RR ($F_3 = dC_2$) and NS ($H_3=dB_2$) three-form field strengths and $F_5=dC_4-\frac{1}{2}C_2\wedge H_3+ \frac{1}{2} B_2\wedge F_3$ the self-dual five-form field strength (for which this action is only a short way of writing the origin of its field equations). The Chern-Simons term is $S_{CS}\propto \int C_4\wedge G_3\wedge \bar{G}_3$. Finally $S_{loc}$ is the contribution from local sources such as D-branes and orientifold planes.

\subsection{Basics of AdS$_5\times$ S$^5$/CFT$_4$ duality}\label{AdS5S5generics}

Let us start recalling some of the relevant results on AdS$_5\times S^5$/CFT$_4$ duality that will be useful to compare with the cases of interest in this article. The original discussion started with the solitonic black brane solutions of the 10D effective action, that has $N$ units of D3-charge; by taking the near horizon limit one extracts the AdS geometry that in the low energy limit can be connected with the world-volume CFT on D3-branes, which is $\mathcal{N}=4$ Yang-Mills in 4D. 

For our purposes, it is more illustrative to approach the AdS$_5$ vacuum from the perspective of flux compactifications of type IIB string theory on $S^5$, since that is the more natural way to compare this background with the KKLT and LVS ones.
One starts in this case from the Freund-Rubin ansatz in which the metric is maximally symmetric, $G_3=0$, the axiodilaton $S$ constant  and  $(F_5)_{mnpqr}\propto \epsilon_{mnpqr}$ (with indices running along the compact dimensions; a similar expression holds for the non-compact dimensions from self-duality of $F_5$). In this way the spacetime is naturally separated in a product of two five-dimensional components. 
 In particular the flux on the compact component, $S^5$ is quantised as:
\be\label{F5flux}
\frac{1}{(2\pi)^4 \alpha'^2}\; \int_{S^5} F_5 = N \:.
\ee

One could try to compactify the 10D theory with a background flux given by \eqref{F5flux}: Plugging the $F_5$ value back into the 10D action and integrating over the five compact extra dimensions and Weyl rescaling to the 5D Einstein frame gives the 5D Einstein-Hilbert term plus a scalar potential for the $S^5$ radius modulus $R_{S^5}$ of the form:
\be
V(R_{S^5})= R_{S^5}^{-16/3} \left(-a+ b N^2 R_{S^5}^{-8}\right) \:.
\ee
The first term comes from the $S^5$ curvature dominating at small $R_{S^5}$ and the second term, dominating at large $R_{S^5}$, comes from the $F_5^2$ term in the action; $a,b$ are  $\mathcal{O}(1)$ positive constants.
Minimising this potential fixes the value of the radius modulus to $R_{S^5}\propto N^{1/4}$. The effective cosmological constant of the non-compact 5D component of the spacetime is given by the value of the potential at the minimum ($\Lambda=V|_{\rm min}$). In this case, it is negative giving rise to AdS$_5$ with  AdS radius equal to the radius of the compact manifold, i.e. $R_{AdS}=R_{S^5}$. This implies that there is no trustable limit in which we can decouple the KK modes. Anyway, this analysis turns out to give the right answer for the background geometry generated by turning on $F_5$ fluxes, as it can be seen by comparing with the solutions of the 10D  equations of motion.
Notice also that the combination of fluxes and curvature of the extra dimensions were enough to fix the overall size of the extra dimensions but there is still a flat direction corresponding to the dilaton which is completely arbitrary. 

To trust the 10D supergravity analysis, one needs to have the AdS radius larger than the string and the 10D Planck scale. This implies that these solutions are valid in the large $N$ and large $g_sN$ limits since\footnote{Notice that from  the second relation we can see that for fixed t'Hooft coupling $\lambda$ the $g_s$ expansion is equivalent to a $1/N$ expansion. Also for fixed $R_{AdS}$ the $\alpha'$ expansion is equivalent to an expansion in  $1/\lambda$.}
\be
\frac{R_{AdS}}{\ell_p^{\rm 10d}}  \sim  N^{1/4} \:, \qquad \qquad
\frac{ R_{AdS}}{\sqrt{\alpha'}} =  \frac{R_{AdS}}{\ell_s} \sim (4\pi g_sN)^{1/4}\equiv \lambda^{1/4} \:.
\ee
At large $N$ and large t'Hooft coupling $\lambda$ the gravity description is well defined whereas for small t'Hooft coupling the perturbative CFT description is well defined.

The symmetries on both sides of the duality match in the sense that local symmetries on the AdS side map to global symmetries on the CFT side. Besides the $\mathcal{N}=4$ supersymmetry, the $SO(4,2)\times SO(6)$ symmetries of the AdS$_5\times S^5$ map to the $SO(4,2)$ 4D conformal symmetry and $SO(6)$  $R$-symmetry of $\mathcal{N}=4$ supersymmetry. The number of degrees of freedom is measured by the `central charge', which is given by $c\sim N^2$. This should be large in order for the duality to work. Also the conformal dimension of different operators has a nontrivial structure. In general, for a scalar particle of mass $m$ the dual CFT$_{d}$ operator has conformal dimension \cite{adscft}
\be\label{ConfDimCFTd}
\Delta=\frac{d}{2}\pm \frac{1}{2}\sqrt{d^2+4 (mR_{AdS})^2}.
\ee
As we discussed before  there is no separation of field theoretical scales since the radius of $S^5$ is the same as $R_{AdS}$. Hence, all Kaluza-Klein (KK) modes have  masses of order $m\sim 1/R_{AdS}$ and therefore there are many operators with conformal dimension of order $\mathcal{O}(1)$.

 \subsection{Calabi-Yau flux compactifications}
 We turn now to phenomenologically interesting Calabi-Yau (CY) flux compactifications that have been shown to be suitable for a controllable moduli stabilisation. 
Without the introduction of extra ingredients, such as background values of p-form potentials,  the simple compactification of type II string theory on such manifolds has plenty of unobserved massless scalars at the 4D EFT level. These scalars are related to the geometric moduli of the Calabi-Yau compact manifold. In type IIB string theory, the relevant ingredients to stabilise the moduli without distorting too much the compact geometry (controlled backreaction) are known: non-zero background values of $G_3$ (three-form fluxes) stabilise the axio-dilaton $S$ and a subset of the geometric moduli, the complex structure moduli $U_\alpha$ ($\alpha=1,...,h^{1,2}$). At lower scales, the rest of the geometric moduli, the K\"ahler moduli $T_i$ ($i=1,...,h^{1,1}$), are stabilised by additional terms in the scalar potential coming from perturbative and non-perturbative $g_s$ and $\alpha'$ corrections.
In this section, we will review the two steps: the first one (GKP) is the same in KKLT and LVS, while they are distinguished by the second one.
 
 \subsubsection*{Axiodilaton and complex structure moduli stabilisation (GKP)}
 Let us give a short review of the relevant features of the Giddings, Kachru, Polchinski (GKP) scenario, in which both complex structure moduli and dilaton are stabilised by switching on three-form fluxes \cite{gkp}.\footnote{See also the previous analogous treatment in the F-theory language, studied in \cite{Dasgupta:1999ss}.} This is at the basis of both  KKLT and  LVS scenarios that we will discuss in the rest of the article.

Compactifying type IIB string theory on a Calabi-Yau orientifold leads to an effective  $\mathcal{N}=1$ supergravity theory in 4D. The low energy action is partially determined by the tree-level K\"ahler potential:
 \be
K= -2\ln \mathcal{V} -\ln i \int\Omega\wedge \Omega^* -\ln (S+S^*)
 \ee
with $\mathcal{V}$ the volume of the Calabi-Yau manifold as a function of the K\"ahler moduli, $\Omega$ the unique $(3,0)$ form as a function of the complex structure moduli and $S=e^{-\phi}+i C_0$ the axiodilaton as before.

The complex structure moduli can be stabilised by turning on RR and NS fluxes $F_3$ and $H_3$, which obey the following quantisation conditions:
 \be\label{FlQuantCondns}
 \frac{1}{(2\pi)^2\alpha'}\int_{\Sigma_A} F_3=M_A \qquad \frac{1}{(2\pi)^2\alpha'}\int_{\Sigma_A} H_3=-K_A \qquad \mbox{with } M_A,K_A \in \mathbb{Z}
 \ee
 for any three-cycles $\Sigma_A\in H_3(X_3)$ of the compact Calabi-Yau three-fold $X_3$. 
 At the level of the 4D effective action  they induce a superpotential \cite{Gukov:1999ya}:
 \be
 W_{flux}=\int G_3\wedge \Omega \:, \qquad \mbox{with} \qquad G_3 = F_3-iS\,H_3 \:.
 \ee
 This superpotential is a function of the  complex structure moduli $U_\alpha$ and dilaton $S$. The supersymmetry conditions $D_\alpha W=D_SW=0$ stabilise their values in terms of the flux numbers $M_A$ and $K_A$ in \eqref{FlQuantCondns}.
\footnote{These conditions are satisfied when the complex structure alligns such that the three-form $G_3$ is imaginary self-dual, i.e. $iG_3=*G_3$. The metric and the five-form $F_5$ are also constrained to depend on a warp factor $e^A$. In particular, the metric on the compact manifold is only conformally equivalent to a Calabi-Yau metric and the compact manifold is called a conformal Calabi-Yau.}  The three-form fluxes $F_3$ and $H_3$ contribute to the effective D3 brane charge. The vanishing of the total D3 brane charge, needed for D3-tadpole cancellation, implies the condition
 \be
\frac{1}{(2\pi)^4{\alpha'}^2} \int F_3\wedge H_3 \, + \,  Q_{D3}^{\rm loc} \,=\, 0 \:,
 \label{tadpole}
 \ee
 where $Q_{D3}^{\rm loc}$ is the contribution coming from the localised sources: D3-branes and supersymmetric gauge fluxes on D7-branes will contribute positively, while O3-planes and curvature of D7-brane and O7-planes contribute negatively (see \cite{gkp}).

The complexified K\"ahler moduli are $T_i=\tau_i+i\vartheta_i$, where $\tau_i$ are the geometric K\"ahler moduli, i.e. the volumes of $h^{1,1}(X_3)$ independent divisors of the Calabi-Yau threefold.
The moduli $T_i$  do not appear in the tree-level superpotential $W_{flux}$  because of the Peccei-Quinn symmetries associated to their axionic component $\vartheta_i$. 
As a consequence, the K\"ahler moduli are flat directions of the tree-level potential, generated by K and $W_{flux}$. In particular the potential is a sum of  positive definite terms, that is minimized at zero by  solving $D_\alpha W=D_S W=0$. 
 
 This situation is similar to the AdS$_5\times S^5$ case in the sense that fluxes stabilise some of the moduli and leave flat directions. In this case the flat directions will naturally be lifted by perturbative and non-perturbative effects in KKLT and LVS.
  
 Varying the values of the integers $K_A,M_A$ generate many different vacua. We may conceive trading the fluxes for D-brane configurations that carry the same information, like described for AdS$_5\times S^5$ at the end of Section \ref{AdS5S5generics}. In this case the configuration would be made up of (p,q) 5-branes wrapping the corresponding three-cycles and being domain walls in the non-compact dimensions. The D3-charge of $F_3,H_3$ would be generated by D3-branes streched between the (p,q) 5-branes. This immediately suggests a `Coulomb branch' approach towards duality. Notice  however that at this stage the spacetime is still Minkowski and not AdS.

 \subsubsection*{KKLT Scenario}
 The KKLT scenario extends the GKP one, adding corrections that allows one to stabilise the K\"ahler moduli. It is assumed that the relevant correction to the scalar potential is a non-perturbative superpotential $W_{\rm np}$ which in general depends on the K\"ahler moduli \cite{Witten:1996bn}:
 \be\label{NPcorrW}
 W_{\rm np}= \sum_i A_i e^{-a_{i} T_i}
 \ee
 with $A_i$ functions of $S,U_\alpha$.  Natural sources of $W_{\rm np}$ are instantonic E3-branes and gaugino condensation effects on the worldvolume of D7-branes, both wrapping four-cycles of the Calabi-Yau manifold. The  assumption of KKLT is that the fluxes can be tuned in such a way that the vacuum expectation value of $W_{flux}$ is $W_{flux}|_{\rm min}\equiv W_0 \sim W_{\rm np}$. Thus the  contributions to $W$ can compete to generate a supersymmetric minimum for the K\"ahler moduli $T_i$, i.e with $D_iW=0$. Consequently, $V\propto -3|W|^2<0$ and so the minimum is AdS$_4$. The vacuum energy gives the value of the cosmological constant, $V|_{\rm min}=\Lambda$. In KKLT we then have (in four dimensional Planck mass $M_p$ units):
 \be\label{KKLTgenericCC}
 \Lambda_{KKLT} \sim - R_{AdS}^{-2}
\sim -\frac{g_s |W_{0}|^2}{\mathcal{V}^2} e^{K_{cs}}\:.
\ee
The $g_s$ factor comes from $e^{K_S}$ with $K_S = -\ln (S+\bar{S} )$. The flux dependent constant $e^{K_{cs}}$ comes from the VEV of the complex structure moduli K\"ahler potential $K_{cs}=-\ln i\int \Omega \wedge \bar{\Omega} $ (where the VEVs depend on the flux numbers). In the following we will absorb this factor in the definition of $W_0$.

Also other scales are fixed (in terms of $M_p$) once we fix all the geometric moduli. The string scale $M_s\sim g_sM_p/\mathcal{V}^{1/2}$ is  larger than the KK scale $M_{KK}\sim g_s M_p/\mathcal{V}^{2/3}$ for volume $\mathcal{V}$ large in string units.  This is similar to the Freund-Rubin cases. However the moduli masses are hierarchically smaller. The complex structure and dilaton masses are of order $m_{S,\alpha}\sim1/\mathcal{V}$. The K\"ahler moduli are even lighter: their masses $m_i\sim |W_{0}|/\mathcal{V}^{1/3}$ are highly suppressed by the exponentially small $|W_{0}|$ factor (with typical values of order $|W_{0}|\sim 10^{-10}$), even if the volume factor is larger than the KK scale (the volume is only parametrically large in KKLT).  

We see that the $|W_{0}|$ factor appears also in \eqref{KKLTgenericCC}.  This implies that there is a hierarchy between the size of the extra dimensions $R_{CY}\sim 1/M_{KK}$ and the AdS radius with ratio:
 \be
 \frac{R_{CY}}{R_{AdS}}\sim \frac{1/M_{KK}}{R_{AdS}}\sim \frac{\mathcal{V}^{2/3}g_s^{-1}}{\mathcal{V}g_s^{-1/2}|W_{0}|^{-1}}\sim \frac{{|W_{0}|}}{\mathcal{V}^{1/3}g_s^{1/2}}\ll 1
 \ee
 This is clearly different from the AdS$_5\times S^5$ case in which both scales are the same. This is important in order to be able to consistently neglect the KK modes in the effective field theory.


Uplifting to de Sitter including supersymmetry breaking was also proposed in KKLT by adding anti-D3-branes. This effect is  under less control and not relevant for the present article. Moreover, the proposed $dS/CFT$ duality is  not that well understood.
 
 \subsubsection*{LVS Scenario}
 The large volume scenario (LVS), also extends GKP but it includes not only the non-perturbative corrections to $W$ \eqref{NPcorrW} but also the perturbative corrections to the K\"ahler potential $K$. In the simplest case the most relevant perturbative contribution is the leading order $\alpha'$ correction which modifies the K\"ahler potential in the following way: 
 \be
 -2\ln \mathcal{V}\rightarrow  -2\ln \left(\mathcal{V}+\xi (S+S^*)^{3/2}\right)
 \ee
 with $\xi$ a constant proportional to the Euler characteristic of the CY. For the generic case of several K\"ahler moduli and $\mathcal{O}(1)$ flux superpotential the K\"ahler moduli are stabilised in such a way that the volume $\mathcal{V}$ is exponentially large. 
In particular, as we will see explicitly in the example studied in Section \ref{SecLVSStab}, the volume $\mathcal{V}$ and another K\"ahler modulus $\tau$ are stabilised such that
\be
\tau\sim 1/g_s> 1 \qquad \mbox{and} \qquad\mathcal{V}\sim e^{a\tau} \gg 1 \:.
\ee
 
Besides the larger value of the volume and the untuned choice of flux superpotential this scenario differs from the KKLT one in several other ways.
The moduli are stabilised at an AdS$_4$ minimum with spontaneously broken supersymmetry. The source of supersymmetry breaking  is the same as in GKP, i.e.  the three-form fluxes: the perturbative and non-perturbative corrections generate only a subleading contribution to the non-zero $D_iW$, where $i$ runs on the K\"ahler moduli.
 The vacuum energy at the minimum goes like 
 \be
 \Lambda_{LVS}\sim -\frac{|W_{0}|^2}{\mathcal{V}^3} g_s^{1/2} e^{K_{cs}} \:.
 \ee
As for KKLT we will absorb the complex structure moduli factor $e^{K_{cs}}$ in $|W_0|^2$.

In LVS  there is a hierarchy of scales but it is  different from that in KKLT. Still  $M_s\sim g_sM_p/\mathcal{V}^{1/2}\gg M_{KK}\sim g_sM_p/\mathcal{V}^{2/3}$ and both are much larger than the gravitino mass $m_{3/2}\sim g_s^{1/2}|W_{0}|M_p/\mathcal{V}$ since the volume is very large $\mathcal{V}\gg 1$. Most moduli masses scale with the volume $\mathcal{V}$ like the gravitino mass, $m_{S,cs,\tau}\sim M_p/\mathcal{V}$, except for the overall volume modulus itself which has a mass of order $m_{\mathcal{V}}\sim  M_p/\mathcal{V}^{3/2}\ll m_{3/2}$ and its axion partner which is essentially massless.\footnote{In the most general cases there may be fields, like those corresponding to K3 fibrations, that get masses only after string loop effects are included and their masses can be smaller than the volume mass $m_f\sim |W_{0}|M_p/\mathcal{V}^{5/3}< m_{\mathcal{V}}$ \cite{cicoli}.}
 
Like in KKLT, also in LVS there is a hierarchy between the CY size and the AdS scale. This hierarchy comes now from having a large volume $\mathcal{V}$ rather than a small flux superpotential $W_{0}$.
\be
 \frac{R_{CY}}{R_{AdS}}\sim \frac{1/M_{KK}}{R_{AdS}}\sim \frac{\mathcal{V}^{2/3}/g_s}{g_s^{1/4}\mathcal{V}^{3/2}/|W_{0}|}\sim \frac{{|W_{0}|}}{g_s^{5/4}\mathcal{V}^{5/6}}\ll 1 \:.
 \ee


\

In Table \ref{TabScales}, we summarise (both for KKLT and LVS) the scales that are relevant for the subsequent sections.

We finally notice that in both KKLT and LVS cases the expansion parameters ($g_s,W_0,\mathcal{V}$) should be related to the exapansion paramenters in the dual CFT, like for the AdS$_5\times S^5$ case where $N$ and $\lambda$ are related to the flux and the string coupling. The difference here is that these parameters  cannot be made arbitrarily small. This is a due to the fact that the flux numbers \eqref{FlQuantCondns} are bounded from above \cite{DenefDouglas1,DenefDouglas2,landscape} by the D3 tadpole cancellation conditions \eqref{tadpole}. This implies on one side that there is a finite number of flux vacua and on the other side that there is a bound on the value of $g_s$ and therefore also on the volume in LVS since $\mathcal{V}\sim e^{a/g_s}$.\footnote{In \cite{landscape} a simple example of a rigid CY is presented. For illustration we use this case to show that $g_s$ will be bounded from below by the tadpole cancellation condition.
For a rigid CY, the flux superpotential is $W_{flux}=(f_1+\Pi f_2)-iS(h_1+\Pi h_2) \equiv F-iSH$, where $\Pi$ is a complex number determined by the geometry. Let us take $\Pi=i$ for simplicity. The susy equation $D_S W_{flux}=0$ gives $\bar{S}=i\frac{F}{H}$. The tadpole cancellation condition is Im$\bar{H}F\leq \mathcal{L}$, where we have separated the D3-brane contribution by the negative contribution coming from O3-planes, D7-branes and O7-planes: $Q_{D3}^{\rm loc}\equiv N_{D3}-\mathcal{L}$. Fixing the S-duality symmetry, the flux vacua satisfying the tadpole cancellation condition are given by $h_2=0$, $0\leq f_1< h_1$ and $\quad h_1f_2\leq  \mathcal{L}$. Thus we have
$
\frac{1}{g_s}\sim \frac{f_2}{h_1}=\frac{h_1f_2}{h_1^2}\leq  \mathcal{L} \:,
$
and hence $g_{smin}\sim \frac{1}{\mathcal{L}}$. (In this computation we are excluding the vacua $h_1=h_2=0$ that would give $g_s=0$, i.e. non-interacting strings.)\label{footnotegsmin}}
This contrasts with the large $N$ expansion in which $1/N$ can be made arbitrarily small.\footnote{We thank N. Seiberg for emphasising this point.}

\begin{table}
\begin{center}
\begin{tabular}{|l|ccccccc|}
\hline 
   & $M_p$     &  $M_s$ & $M_{KK}$ & $R_{AdS}^{-1}$ & $m_{S,\alpha}$ & $m_{i\neq \mathcal{V}}$ & $m_{\mathcal{V}}$\\ \hline
KKLT & $1$ & $\frac{g_s}{\mathcal{V}^{1/2}}$ & $\frac{g_s}{\mathcal{V}^{2/3}}$ & $\frac{g_s^{1/2}|W_0|}{\mathcal{V}}$ & $\frac{1}{\mathcal{V}}$ & $\frac{|W_0|}{\mathcal{V}}$ & $\frac{|W_0|}{\mathcal{V}}$ \\
LVS & $1$ & $\frac{g_s}{\mathcal{V}^{1/2}}$ & $\frac{g_s}{\mathcal{V}^{2/3}}$ & $\frac{g_s^{1/4}|W_0|}{\mathcal{V}^{3/2}}$  & $\frac{1}{\mathcal{V}}$ & $\frac{|W_0|}{\mathcal{V}}$ & $\frac{|W_0|}{\mathcal{V}^{3/2}}$ \\ \hline
\end{tabular}
\caption{Relevant scales of KKLT and LVS scenario, in 4D Planck units: string scale, KK scale, AdS scale, axiodilaton and complex structure moduli masses, K\"ahler moduli masses, volume modulus mass.}\label{TabScales}
\end{center}
\end{table}

\section{Properties of the \boldmath CFT$_3$ duals}\label{sec:PropertiesCFT3}
 
Having a precise description of the AdS$_4$ type IIB flux vacua,  it is natural to search for the CFT$_3$ duals. The situation is much less clear than in the AdS$_5\times S^5$/CFT$_4$ case. The main obstacle is that there is no clean 10D string theory formulation of the KKLT and LVS scenarios, and most of the results are obtained only through an effective field theory approach. In particular, the description of the non-perturbative effects is valid only within the effective field theory approximation. Contrary to the AdS$_5\times S^5$ case there are no known black-brane solutions in which the AdS factor can be achieved by a near horizon limit.
On the other hand we should be able to extract some partial information based on the effective field theory results and by analogy with known cases.

  In particular the study of the Coulomb branch motivated \cite{eva} to come-up with a concrete proposal for the duals of KKLT compactifications. As anticipated before, the main idea is to consider (p,q) 5-branes  that are domain walls separating AdS vacua corresponding to different fluxes. These 4D  domain walls are 5-branes wrapping the same 3-cycles threaded by the fluxes and located at different points in the radial direction of AdS.\footnote{Notice that these are precisely the same brane configurations that can nucleate the potential decay of metastable minima as discussed in \cite{stability}.} D3-branes  must be introduced in order to satisfy  the total D3 charge constraint (\ref{tadpole}). These D3 branes will be stretched between the 5-branes. As for the AdS$_5\times S^5$ case, the domain wall configurations should represent the dual CFT in its Coulomb branch, i.e. when the fields representing the location of the corresponding branes get a non-zero VEV. This is an interesting proposal that is analogous to the AdS$_5\times S^5$ case: it implements a brane/flux duality that seems to be at the core of the gauge/gravity correspondence. However it is not yet clear if this is the proper identification of the CFT.
 
In general, the understanding of the CFT side is very limited. Hence, rather than concentrating on tests of the duality, we will focus on extracting properties that these CFTs will have in order to be dual to the KKLT or LVS AdS$_4$ minima.
In reference \cite{papa2} a set of conditions were spelled out in order for a CFT to have a gravity dual: (i) Having a large central charge $c$; (ii) A small set of operators of conformal dimension of 
$\mathcal{O}(1)$ and (iii) approximate (in an $1/\sqrt{c}$ expansion) factorisation of their correlation functions. In the following we will see that if a CFT dual exists that is dual to KKLT or LVS AdS minima, then it will satisfy the properties just mentioned.

\subsection{Central charge and number of degrees of freedom}  
In 2+1 dimensional CFTs the central charge ($c\sim  N_{\rm dof}$ ) can be defined at least in two  ways \cite{cardy}: from the two point function of the energy momentum tensor or from the `entropy/temperature relation'. Both definitions were proven to be equivalent for theories with AdS duals \cite{adam} and to be proportional to $R_{AdS}^2$ in 4D Planck units. So we can write:
\be
N_{\rm dof}\sim R_{AdS}^2  \sim  \l\{  \begin{array}{lcl}  \frac{\mathcal{V}^2}{g_s|W_{0}|^2} &\qquad & \rm{KKLT}\\  \\ \frac{\mathcal{V}^3}{g_s^{1/2}|W_{0}|^2} &\qquad & \rm{LVS} \\ \end{array} \r.
\ee
We see that in both cases, KKLT and LVS, the CFT has a very large central charge, as expected for a CFT that has a gravity dual. This should be interpreted as the analogue of large $N$.\footnote{Comparing to the $AdS_5\times S^5$ case the central charge is the natural generalisation of the number of colors $N$ (since in that case $c\sim N^2$). However there is no  clear analogue of the 't Hooft coupling $\lambda$. In any case, we may assume the relation $\lambda\sim g_s N$ suggested by the Riemann surface topologies that organise the 't Hooft and string theory expansions. Hence, we may identify a 't Hooft-like coupling as $\lambda\sim g_s N_{\rm dof}^{1/2}$ with $N_{\rm dof}$ as above. We thank the referee for this suggestion.}


The number of degrees of freedom should match with the one computed in the dual CFT. If one consider the ensemble of flux vacua, there will be a vacuum with the smallest cosmological constant, i.e. the vacuum with the maximum number of degrees of freedom $N_{\rm dof}^{\rm max}$. If one knows the distribution of $\Lambda$ over the Landscape of flux vacua and the total number of vacua $\mathcal{N}_{\rm vac}$, one can estimate what is the minimum value that the cosmological constant will take in the Landscape. For KKLT this problem was studied in \cite{eva}: Expressing the volume in terms of the flux dependent parameters $g_s,W_{0},A$ and knowing that the distributions of such quantities are roughly uniform, one obtains a roughly uniform distribution of $\Lambda$ \cite{DenefDouglas1,DenefDouglas2}. This means that $\Lambda_{\rm min}^{KKLT}\sim \frac{1}{\mathcal{N}_{\rm vac}}$ and so $N_{\rm dof}^{\rm max,KKLT}=\mathcal{N}_{\rm vac}$.

In the LVS case, the value of $\Lambda$ at the minimum is given by $\Lambda^{LVS}\sim \frac{A^3 e^{-3a/g_s}}{|W_{0}|}$ (where we have used $\mathcal{V}\sim \frac{W_0}{A}e^{a/g_s}$). Because of the exponential factor, the distribution will be extremely peaked at small values of $\Lambda$ (see \cite{w0bound} for a recent discussion of this point). This leads to the expectation that the minimal value of $\Lambda^{LVS}$ will be much smaller than the minimal value of $\Lambda^{KKLT}$. Because of the exponential relation between $\Lambda^{LVS}$ and $g_s$, the smallest value of $\Lambda^{LVS}$ over the space of flux vacua is realised when $g_s$ takes the minimal value (and $W_0$ is of order one). 

One may try to estimate the minimal value of $g_s$ by considering its uniform distribution around zero and making analogous consideration as for $\Lambda_{KKLT}$. Unfortunately, $g_s\sim 0$ is at the bounday of the moduli space and one needs to be careful. Moreover, the uniform distribution is valid up to the value of $g_s$ for which the continuous approximation is valid. In \cite{DenefDouglas1} this bound was computed for the rigid Calab-Yau case: the continuous approximation is valid for $g_s \geq \frac{1}{\sqrt{\mathcal{L}}}$, where $\mathcal{L}$ is the D3-charge of the localised sources. This bound is quite big, compared to $\frac{1}{\mathcal{N}_{\rm vac}}$ (that is the minimal $g_s$ that would be estimated if the continuous uniform distribution were valid for all values of $g_s$), that for this case is equal to $\frac{1}{\mathcal{L}^2}$ \cite{DenefDouglas1}. On the other hand, this does not mean that there are not flux vacua realising $g_s \leq \frac{1}{\sqrt{\mathcal{L}}}$. In fact, as shown in footnote \ref{footnotegsmin}, the actual minimal number of $g_s$ is $\frac{1}{\mathcal{L}}$.
For the generic case, it is difficult to estimate how small $g_s$ can be without the continuous approximation. Moreover, this bound is valid for the rigid CY, i.e. with $h^{1,2}=0$. For CYs with large $h^{1,2}$, one expects that this bound is consistently lowered, even though maybe not at the level of $\frac{1}{\mathcal{N}_{\rm vac}}$.

We can anyway try to infer at least the relation $\Lambda_{\rm min}^{LVS} < \Lambda_{\rm min}^{KKLT}=\frac{1}{\mathcal{N}_{\rm vac}}$. Considering $W_0,A,a\sim \mathcal{O}(1)$, this condition becomes $g_s < \frac{1}{\ln \mathcal{N}_{\rm vac}}$. In a situation with many flux vacua, it is not hard to believe that this inequality is satisfied as there is a large set of tunable fluxes that can make $g_s$ to be much smaller than $\frac{1}{\ln \mathcal{N}_{\rm vac}}$. We checked this in an example published in \cite{MartinezPedrera:2012rs}\footnote{We thank M.~Rummel for providing the unpublished results concerning such an example.}. In that article, the authors studied  Type IIB compactification on $\mathbb{CP}^4_{11169}[18]$ (the same CY we used for our analysis at the end of Section \ref{SecLVSStab}), with only a subset of flux vacua turned on (see \cite{Denef:2004dm} for an explicit treatment). These fluxes were anyway enough to stabilise all the complex structure and the dilaton (at a symmetric point in the complex structure moduli space). The number of flux vacua after moduli stabilisation is $\mathcal{N}_{\rm vac}\sim 10^{12}$, while $\mathcal{L}\sim \mathcal{O}(100)$ \cite{Denef:2004dm}. In \cite{MartinezPedrera:2012rs} the authors were able to explicitly scan only a subset of such flux vacua, i.e. $\mathcal{O}(10^4)$ vacua. Among these, they found that the minimal value of $g_s$ is $g_s^{\rm min}=\frac{1}{27152}$.  We see that this value is much smaller than $\frac{1}{\ln \mathcal{N}_{\rm vac}}\sim \frac{1}{30}$. We expect that if it was possible to compute $g_s$ for all the $10^{12}$ flux vacua in the considered subset, the actual minimal value of $g_s$ could even be lowered.\footnote{If computer techniques will be improved in the next future, a complete scan of flux vacua can be studied (including all the bulk three-form fluxes and the two-form fluxes on the D7-branes), enlarging the number of $\mathcal{N}_{\rm vac}$ to the famous $10^{500}$ (or even $10^{2000}$ if one includes the D7-brane fluxes) \cite{Denef:2004dm}  and correspondingly being able to probe much smaller values of $g_s$.
}
Hence we can conclude that in this example $\Lambda_{\rm min}^{LVS} \ll \Lambda_{\rm min}^{KKLT}$, as we guessed by considerations on the distribution of the cosmological constant in the two setups. The  example we have considered is typical in the landscape of type IIB compactifications and the conclusion can be generalised to other Calabi-Yau manifolds. 

To summarise, in this section we have argued that the number of degrees of freedom in the dual CFT is very different for KKLT and LVS. In particular the maximal value that $N_{\rm dof}$ can take (given by the minimal value of $\Lambda$) is much bigger for LVS with respect to the one for KKLT. We do not have a clear interpretation why this happens. Without a complete scan of flux vacua in concrete type IIB compactifications (that is really hard to do with the present techniques and not the main point of this article), we are not able to estimate how huge the number of degrees of freedom is for LVS.

\subsection{Conformal dimensions}
The relation between the mass $(m)$ of the various fields on the gravity side and the conformal dimension ($\Delta$) of the operator in the dual CFT is given in \eqref{ConfDimCFTd} for scalar fields. In our case ($d=3$):
\be\label{DeltaExpr}
m^2R_{AdS}^2=\Delta(\Delta-3) \:.
\ee

\begin{itemize}
\item{} {\bf KKLT}:
Since there is a hierarchy of scales we know that the conformal dimensions of string and KK modes will be hierarchicaly large. The relevant fields are the moduli. The complex structure moduli and dilaton have masses of order $\sim 1/\mathcal{V}$ whereas the K\"ahler moduli have masses of order the gravitino mass $m\sim m_{3/2}\sim |W_{0}|/\mathcal{V}$. Therefore, from \eqref{DeltaExpr} we have 
\be
\Delta_{moduli}\sim  \l\{ \begin{array}{lcl} \mathcal{O}(1) &\qquad & \tau_i ,\, \vartheta_i \,, \\ \frac{1}{|W_{0}|} \gg 1 & \qquad & U_\alpha ,\, S\,, \end{array}\r.
\ee
where $T_i=\tau_i+i \vartheta_i$ are the K\"ahler moduli, $U_\alpha$ the complex structure moduli and $S$ the axiodilaton.
For a typical CY there is a relatively large but finite number ($h^{1,1} \sim \mathcal{O}(1-100)$) of fields with $\mathcal{O}(1)$ conformal dimension. Since there is a gravity dual we expect approximate factorisation of the correlation functions for these operators.
\\

\item{} {\bf LVS}:
The masses of the various moduli go as 
\ben
&&m_{\tau_s}\sim m_{a_s}\sim m_{3/2}\sim \frac{|W_{0}|}{\mathcal V}\:,\nn\\
&&  m_U\sim m_S\sim \frac{1}{\mathcal V}\:,\nn\\
&& m_{\tau_b}\sim\frac{|W_{0}|}{{\mathcal V}^{3/2}}\:, \nn\\
&&m_{\vartheta_b}\sim 0 \:, \nn
\een
where we have omitted the irrelevant $g_s$ factors and we are taking a model with one large ($\tau_b$) and one small ($\tau_s$) K\"ahler modulus. 
From these expressions we get that
\ben
&&m^2_{\tau_s}R^2_{AdS}\sim m^2_{\vartheta_s}R^2_{AdS}\sim {\mathcal V}\gg 1 \:,\nn\\
&& m^2_UR^2_{AdS}\sim m^2_SR^2_{AdS}\sim {\mathcal V} \gg 1\:,\nn\\
&& m^2_{\tau_b}R^2_{AdS}\sim {\mathcal O}(1)\:, \nn \\
&&m^2_{\vartheta_b}R^2_{AdS}\sim 0 \:.
\een
The above equations suggest that the conformal dimension of the operators dual to complex structure and small K\"ahler moduli is very large whereas for the operators dual to volume modulus ($\mathcal{V}\sim \tau_b^{3/2}$) and its axionic partner it is ${\mathcal O}(1)$:
\be
\Delta_{moduli}\sim  \l\{ \begin{array}{lcl} \mathcal{O}(1) &\qquad & \tau_b,\vartheta_b\,, \\ \mathcal{V}^{1/2}  \gg 1 & \qquad & \tau_s,\, \vartheta_s ,\, U_\alpha ,\, S\:.  \end{array}\r.
\ee
Since there are only few operators with ${\mathcal O}(1)$ conformal dimension, it suggests that the dual field theory is very strongly coupled. Again correlation functions should approximately factorise.

We find this result particularly interesting since the CFT seems to have only one scalar operator (and its axionic partner) with conformal dimension of $\mathcal{O}(1)$. This is related to the fact that the volume modulus mass is hierarchically smaller than the gravitino mass, despite supersymmetry being broken. A standard concern about this result is if quantum effects, after supersymmetry breaking, will naturally raise the value of this mass to the supersymmetry breaking scale. This issue was discussed in \cite{uber} in which the loop corrections to the modulus masses were found to be proportional to $\delta m^2\propto g\cdot \Delta m_{\rm bos-ferm}^2 \sim \frac{m_{KK}^2}{M_p^2}m_{3/2}^2 \sim \frac{M_p^2}{{\cal V}^{10/3}}$. We see that for very large volume $\mathcal{V}$, $\delta m \ll m_{\tau_b} \ll m_{3/2}$. It is then expected that in the corresponding CFT quantum corrections will not substantially change the conformal dimension and keep this hierarchy. Having a CFT with such a simple structure of low-lying operators is intriguing and may be interesting to search for.

\end{itemize}

\subsection{Wrapped branes and their dual}

There are some operators in the dual field theory whose existence depends on the given choice of flux vacuum. This allows us to distinguish two different flux vacua that have the same value of $W_0$, $g_s$ and $A$. One such class of operator we consider here is the baryon like operator/vertex. These operators/vertices in the field theory are dual to the configuration of Dp-brane wrapping p-cycle in compact directions. They have provided non trivial checks of AdS/CFT duality \cite{Witten:1998xy,Gubser:1998fp}. In our case it is very natural to consider a configuration of D3-branes wrapping a three-cycle $\Sigma$ of the  CY manifold. This will correspond to a massive particle in AdS$_4$ whose mass is determined in terms of the volume of the three-cycle. Assuming that the particle is stable, we want to find the operator or vertex in the CFT dual. 

On the D3-brane world volume there is a gauge field $A_\mu$. 
The D3-brane Chern-Simons action generates a coupling between this gauge field and the background fluxes (in the combination involving the RR scalar field):
\be
(2\pi)\alpha'\mu_3\int_{\Sigma\times \mathbb R}A\wedge\left[F_{(3)}+C_0H_{(3)}\right].
\ee
Here $\mu_3=\frac{1}{(2\pi)^3\alpha'^2}$ is the D3-brane charge.
Now, using \eqref{FlQuantCondns} 
we find that the background fluxes contribute to the charge of the particle under the worldvolume $U(1)$  symmetry, which is given~by
\be
\left[M_{\Sigma}-C_0K_{\Sigma}\right]\int_{\mathbb R}A \:.
\ee
The charge $\left[M_{\Sigma}-C_0K_{\Sigma}\right]$ must be cancelled in order to prevent a tadpole for the field $A$. Hence there must be an opposite contribution coming from open strings attached to the D3-brane. The charge coming from open string ends is integral and so it can cancel the one generated by fluxes only if the last one is integral as well. While $M_{\Sigma}$ and $K_{\Sigma}$ must be integral,  $C_0$ is a not necessarily an integer depending on fluxes (after  moduli stabilisation). We conclude that the condition for the baryon operator to be present in the dual theory is that 
\begin{equation}
\left[M_{\Sigma}-C_0K_{\Sigma}\right]\in \mathbb Z \:.
\end{equation}

Different choice of fluxes that give the same value of $W_0$ and $g_s$ can allow different operators in the dual theory. These operators are a useful ingredient to probe different flux vacua.



\section{Effective potential and quantum logarithmic effects}\label{sec:EffPot}

In AdS/CFT duality, the partition function of the theory of gravity on AdS space is equal to the partition function of the CFT living at it's boundary \cite{Gubser:1998bc,Witten:1998qj}. There have been several checks for this duality but the majority of works are in the infinite N limit which corresponds to studying classical gravity in the bulk.

In this section we briefly review how one-loop corrections to the partition function in gravity systems have been used to learn and test the gauge/gravity duality. Loop corrections to the partition function on the gravity side correspond to going beyond planar limit on the dual field theory side. This provides a test of AdS/CFT duality beyond planar limit which is very non-trivial, as it involves string loop computations on the AdS side. However in the $\alpha'\rightarrow 0$ limit, this  reduces to the computation in supergravity. In these procedure it is always worth looking for the quantity which does not depend on the details of the UV theory. Such quantities are universal in the sense that they can be calculated in the low energy effective field theory. 
One quantity of this type is the logarithmic correction, $\ln (R_{AdS})$, in the partition function of effective field theory on the gravity side. This object has been used quite successfully in studying the entropy of black holes \cite{Banerjee:2010qc,Banerjee:2011jp,Sen:2011ba,Gupta:2014hxa,Chowdhury:2014lza}:  the logarithmic corrections calculated on the supergravity side matches with those computed on the string theory side. A similar comparison has been made in \cite{Bhattacharyya:2012ye} where the supergravity calculation in AdS$_4\times X_7$, where $X_7$ is a compact seven dimensional manifold, reproduces the correct coefficient of the logarithmic correction present in the $\frac{1}{N}$-expansion of the partition function of the three dimensional CFT. 

Motivated by this success, we will do a similar computation in the KKLT and the LVS cases where we have supersymmetric and non supersymmetric AdS$_4$ minima respectively. Assuming the validity of the AdS$_{d+1}$/CFT$_{d}$ duality, these vacua will have a dual description in terms of a (unknown) three-dimensional CFT. The computation on the AdS side will give a non trivial prediction for the CFT partition function. As we will explain in detail below, the logarithmic correction, $\ln(R_{AdS}\epsilon)$, arises at one loop when a particle whose mass scales with some power of $R_{AdS}$ runs inside the loop. Calculating such logarithmic corrections in KKLT and LVS requires the knowledge of the explicit form of masses of all the moduli fields.  These are not available at the moment for all the scalar fields. In particular, for the compactifications we have considered, the K\"ahler moduli masses are known as functions of few paramenters (depending on the flux numbers \eqref{FlQuantCondns}), while the complex structure moduli masses are unknown functions of the fluxes. 
Since all the masses of the  K\"ahler moduli and gravity multiplets scale with some power of $W_0$ (a function of the flux numbers), we will calculate a similar logarithmic correction, $\ln |W_0|^2$, that does not requires the knowledge of the complex structure moduli masses (that do not scale with $W_0$).  We claim that this is a universal prediction for the dual CFT, once one identifies what $W_0$ parametrises in the dual theory.

\subsection{The limit  $|W_0|\rightarrow 0$}
The effective field theory in KKLT and LVS (after integrating out the axiodilaton and the complex structure moduli) are basically labelled by three parameters, that are functions of the flux numbers \eqref{FlQuantCondns}: the super potential $W_0$,  string coupling $g_s$ and the prefactor $A$ of the non-perturbative contribution to the superpotential (in case there is only one non-negligible non-perturbative effect). After stabilising the K\"ahler moduli, these fields are also function of these parameters. In particular, this happens for the volume of the compactification manifold $\mathcal{V}=\mathcal{V}(W_0,g_s,A)$. Inverting this relation, we can express $A$ in terms of $\mathcal{V}$ and use this last one as the third parameter. The radius of the AdS is given by
\be
\frac{1}{R_{AdS}^2}\sim \frac{g^\alpha_s W_0^2}{\mathcal V^\beta} \:.
\ee
Here $\alpha=1, \beta=2$ for KKLT and $\alpha=\frac12, \beta=3$ for LVS.

Now, in order for the supergravity approximation to work, $R_{AdS}$ needs to be arbitrarily large. This limit can be achieved in various ways. However in our case we will work in the limit,
\be\label{limitW0}
W_0\rightarrow 0,\quad g_s=\text{fixed} ,\quad \mathcal V=\text {large but fixed} \:.
\ee 
We motivate this as follows: if we are interested in the coefficient of logarithmic correction $\ln R_{AdS}$, which is the general quantity of interest in standard AdS/CFT duality, then we need to include all the fields whose mass scales with some power of $R_{AdS}$. Therefore in order to calculate logarithmic correction $\ln R_{AdS}$, we need to know the masses of all the moduli fields including the KK modes. This is a rather harder problem at present, due to the unknown expression for the complex structure moduli masses.
An important point to observe is that the masses of KK modes and complex structure moduli do not scale with $W_0$, while the masses of the K\"ahler moduli, gravitino mass and the cosmological constant scale do scale with $W_0$. Hence only K\"ahler moduli and the gravity multiplet contribute to the coefficient of $\ln |W_0|^2$, and we can single this out by considering the limit \eqref{limitW0}. This is the reason why we look for the coefficient of $\ln |W_0|^2$.

\subsection{Effective potential\label{sub:Effective-potential}}

In this section we will describe the computation of the one loop effective action in supergravity coupled to matter fields.  The one loop calculation involves the computation of determinants of the various operators which appear at the quadratic order in the fluctuations of the fields in the Lagrangian about the background fields.  The determinants are then expressed in terms of the heat kernel of the operator. The UV divergences of the effective action is captured by the asymptotic expansion of the heat kernel. In this expansion we will look for the logarithmic divergence.

The heat kernel expression for the one-loop effective action is:
\begin{align}
\Gamma_{(1)} & =-\frac{1}{2}\int_{\epsilon}^{\infty}\frac{d\tau}{\tau}{\rm Str}\exp[-\tau(\boldsymbol{\nabla}^{2}+\boldsymbol{X}+{\cal \boldsymbol{M}}^{2})]\nonumber \\
= & -\frac{1}{2}\int_{\epsilon}^{\infty}\frac{d\tau}{\tau}{\rm Str}\{\exp[-\tau(\boldsymbol{\nabla}^{2}+\boldsymbol{X})]e^{-\tau{\cal \boldsymbol{M}}^{2}}\}.\label{eq:Gamma1}
\end{align}
Here $\boldsymbol{\nabla}^{2}=\boldsymbol{-I}g^{\mu\nu}\nabla_{\mu}\nabla_{\nu}$
where $\boldsymbol{I}$ is the unit matrix in the space of fields
and $\boldsymbol{X}$ is a spin dependent matrix that is linear in
the Riemann tensor \cite{Christensen:1978md} (the gauge field background
in 4D has been taken to be zero) and $\boldsymbol{M}$ is a field dependent
mass matrix. In the second line we have dropped space time derivatives
of $\boldsymbol{M}$ since we are just considering the effective potential.
Now we use the adiabatic expansion for the heat kernel to write
\begin{align}
\Gamma_{(1)} & =-\frac{1}{2}\int_{\epsilon}^{\infty}\frac{d\tau}{\tau}\frac{1}{16\pi^{2}\tau^{2}}{\rm STr}\{[\boldsymbol{a}_{0}^{({\rm s})}\boldsymbol{I}+\boldsymbol{a}_{2}^{({\rm s})}\tau+\boldsymbol{a}_{4}^{({\rm s})}\tau^{2}+\ldots]e^{-\tau{\cal \boldsymbol{M}}^{2}}\}\nonumber \\
 & =-\frac{1}{32\pi^{2}}{\rm Str}[\boldsymbol{a}_{0}^{({\rm s})}\boldsymbol{I}_0+\boldsymbol{a}_{2}^{({\rm s})}\boldsymbol{I}_2+\boldsymbol{a}_{4}^{({\rm s})}\boldsymbol{I}_{4}+\ldots] \:.\label{eq:GammAdiabatic}
\end{align}
Note that in the first line above the trace includes an integral over the
space time. Also the prefix `$S$' on the trace simply implies tracing
over the physical degrees of freedom with a factor $(-1)^{2{\rm s}}$, ${\rm s}$
being the spin. The coefficients $\boldsymbol{a}_{2n}^{({\rm s})}$ are integrals over the
De Witt coefficients and are given below \cite{ Christensen:1978md,Vassilevich:2003xt}:
\begin{align}
\boldsymbol{a}_{0}^{({\rm s})} & =\int d^{4}x\sqrt{g}Tr \boldsymbol{I} \:,\nn\\
\boldsymbol{a}_{2}^{({\rm s})} & =\frac{1}{6}\int d^{4}x\sqrt{g}Tr(R+6X^{{\rm s}})\:,\label{eq:DW4}\\
\boldsymbol{a}_{4}^{({\rm s})} & =\frac{1}{180}\int d^{4}x\sqrt{g}\{\alpha^{{\rm s}}C_{\mu\rho\nu\sigma}C^{\mu\rho\nu\sigma}+\beta^{{\rm s}}(R_{\mu\nu}R^{\mu\nu}-\frac{1}{4}R^{2})+\gamma^{{\rm s}}\square R+d^{{\rm s}}R^{2}\}\:.\nn
\end{align}
Here $Tr$ indicates the trace over the various indices of the field like space time indices and internal indices. 

In \eqref{eq:GammAdiabatic} $\boldsymbol{I}$ is a unit matrix and $\boldsymbol{I}_{0,2,4}$
are matrix valued integrals, whose entries are of the form\footnote{Due to the UV divergence, we need to use a cutoff $\epsilon$.  In string theory $\epsilon$ is a physical
cutoff, effectively $\epsilon=l_{s}^{2}$ or $l_{KK}^{2}$.}
\[
I_{0}=\int_{\epsilon}^{\infty}\frac{d\tau}{\tau^{3}}e^{-\tau m^{2}},\, I_{2}=\int_{\epsilon}^{\infty}\frac{d\tau}{\tau^{2}}e^{-\tau m^{2}},\, I_{4}=\int_{\epsilon}^{\infty}\frac{d\tau}{\tau}e^{-\tau m^{2}}\:.
\]
These integrals satisfy the conditions
\[
\frac{dI_{2}}{dm^{2}}=-I_{4},\,\frac{dI_{0}}{dm^{2}}=-I_{2}.
\]
Finally we have (substituting $t=m^{2}\tau$) 
\[
I_{4}=\int_{\epsilon m^{2}}^{\infty}\frac{dt}{t}e^{-t}=\Gamma(0,\epsilon m^{2}) \:,
\]
where 
\[
\Gamma(z,x)\equiv\int_{\epsilon}^{\infty}t^{z-1}e^{-t}
\]
 is the incomplete Gamma function for which we have the expansion
(for $z=0$),
\[
\Gamma(0,x)=-\gamma-\ln x-\sum_{k=1}^{\infty}\frac{(-x)^{k}}{k(k!)}.
\]
Thus $I_{4}=-\ln(\epsilon m^{2})+f(\epsilon m^{2}),\, I_{2}=m^{2}\ln(\epsilon m^{2})+m^{2}g(\epsilon m^{2}),\, I_{0}=-1/2 m^{4}\ln(\epsilon m^{2})+m^{4}h(\epsilon m^{2})$
where $f$ is an analytic function and $g,h$ are meromorphic functions
with poles of order 1 and 2 respectively. Putting these results into \eqref{eq:GammAdiabatic}, we obtain
\begin{align}
\Gamma_{(1)} & =\frac{1}{32\pi^{2}}{\rm Str}\left[\frac{1}{2}\boldsymbol{a}^{({\rm s})}_{0}\boldsymbol{M}^{4}-\boldsymbol{a}^{({\rm s})}_{2}\boldsymbol{M}^{2}+\boldsymbol{a}^{({\rm s})}_{4}\right]\ln(\epsilon\boldsymbol{M}^{2})\nonumber \\
 & +\int d^{4}x\sqrt{g}\left[\frac{1}{\epsilon}{\rm Str}\boldsymbol{M}^{2}+\frac{1}{\epsilon^{2}}{\rm Str}\boldsymbol{I}+V_{(1)}(\epsilon\boldsymbol{M}^{2})\right].\label{eq:Gamma1-1}\\
\nonumber 
\end{align}
with the last integrand $V_{(1)}$ being an analytic function.
In a theory with equal numbers of fermionic and bosonic degrees of
freedom such as a supersymmetric theory the $\epsilon^{-2}$ term
will vanish. In a supersymmetric theory with zero cosmological constant and unbroken
supersymmetry the $O(\epsilon^{-1})$ will also vanish. In a flat
background we will only have the first term in the factor multiplying
$\ln(\epsilon\boldsymbol{M}^{2})$ which of course gives the usual
Coleman-Weinberg formula. In the following we will focus on the log
divergence term, the first line in \eqref{eq:Gamma1-1}, since the coefficients are independent of the UV regulator and we can find a universal quantity that is just proportional to the log of the flux superpotential.\footnote{Note that modes with masses close to the cutoff, like KK modes and string states, give a suppressed contribution to the first line of \eqref{eq:Gamma1-1}. In any case as noted earlier these will not contribute to the $\ln W_0$ terms.}

\subsection{Effective potential \boldmath $\Gamma_{(1)}$ about AdS background}

We now compute the De Witt coefficients $\boldsymbol{a}_i$ appearing in the logarithmic divergence for the fields with spin$\leq 2$ about the AdS$_4$ background. In the next section we will use these coefficients to compute the $\ln |W_0|$ term for the cases of KKLT and LVS flux compactifictions.

The AdS$_4$ metric is given by
\be
ds^2=R_{AdS}^2\l(d\eta^2+\sinh^2\eta d\Omega_3\r),
\ee
where $d\Omega_3$ is the metric of three-sphere.\footnote{The metric is given in Poincare coordinates by
$
ds^2=R_{AdS}^2\frac{1}{z^2} (dz^2+\sum_{i=1}^{3}(dx^i)^2)
$.
This presentation shows  that AdS is conformally flat so that its Weyl tensor is manifestly equal to zero, $C_{\mu\rho\nu\sigma}=0$.}
In this background the curvature has the form 
\begin{equation}
R_{\mu\rho\nu\sigma}=-\frac{1}{3}L^{-2}(g_{\mu\nu}g_{\rho\sigma}-g_{\mu\sigma}g_{\nu\rho}),\, R_{\mu\nu}=-L^{-2}g_{\mu\nu},\, R=-4L^{-2}\:,\label{eq:Rads}
\end{equation}
where $R_{AdS}^{2}=3L^{2}$ and  $-L^{-2}\equiv -|\Lambda |<0$ is the AdS cosmological constant (CC). 

Let us evaluate the coefficient
$\boldsymbol{a}^{({\rm s})}_{4},\boldsymbol{a}^{({\rm s})}_{2}$ and $\boldsymbol{a}^{({\rm s})}_{0}$ in this background. From \eqref{eq:Rads} we have $R^{2}=16L^{-4},\, R_{\mu\nu}R^{\mu\nu}=4L^{-4}$,
and thus for our background we also have
\be
R_{\mu\nu}R^{\mu\nu}-\frac{1}{4}R^2=0,\quad C_{\mu\rho\nu\sigma}C^{\mu\rho\nu\sigma}=0 \:.
\ee
 We parametrize the De Witt coefficients (\ref{eq:DW4}) as follows,
\be
\boldsymbol{a}_{4}^{({\rm s})}=\frac{d^{{\rm s}}}{180}\int d^{4}x\sqrt{g}R^2,\quad 
\boldsymbol{a}_{2}^{({\rm s})}=\frac{c^{{\rm s}}}{6}\int d^4x\sqrt{g}R,\nn
\ee
\be\label{GilkCoef1}
\boldsymbol{a}_{0}^{({\rm s})}=f^{\rm s}\int d^4x\sqrt{g} \:.
\ee
The coefficients ($d^{{\rm s}},c^{{\rm s}},f^{{\rm s}}$) are given in Table 1 (for details see Appendix \ref{AppB}). Suppose the theory has neutral chiral supermultiplets (moduli/ini)
along with the graviton and the gravitino.
Then we have the effective potential\footnote{Note that we have suppressed for simplicity an additional sum over chiral scalar multiplets - this will be remedied later.},
\ben
\Gamma_{(1)}& \sim&\frac{1}{32\pi^2}\sum_{{\rm s}=0}^2(-1)^{2{\rm s}}\l[\frac{1}{2}\boldsymbol{a}_0^{\rm s} m_{\rm s}^4-\boldsymbol{a}_2^{\rm s} m_{\rm s}^2+\boldsymbol{a}_4^{\rm s}\r]\ln(\varepsilon m_{\rm s}^2)\nn\\ &\sim&\frac{1}{32\pi^2}\int\sqrt{g}d^{4}x\sum_{s=0}^2(-1)^{2s}\l[\frac{1}{2}f^sm_s^4-\frac{c^s}{6}Rm_s^2+\frac{d^s}{180}R^2\r]\ln(\varepsilon m_s^2)\:,
\een
where we have used (\ref{GilkCoef1}).

The volume of AdS$_4$ is infinite, however in AdS/CFT there is a well defined prescription to extract the finite part \cite{Diaz:2007an},
\be
\int d^4x\sqrt{g}=\frac{4\pi^2R_{AdS}^4}{3} =12\pi^2 L^4 \:.
\ee
Thus we get,
\be\label{CoefflnEm2Gen}
\Gamma_{(1)} \sim\frac{3 L^4}{8}\sum_{{\rm s}=0}^2(-1)^{2{\rm s}}\l[\frac{1}{2}f^{\rm s} m_{\rm s}^4+\frac{2c^{\rm s}}{3L^2}m_{\rm s}^2+\frac{4d^{\rm s}}{45L^4}\r]\ln(\varepsilon m_{\rm s}^2)\:.
\ee
While carrying out the above computations, we also need to include the contributions of the various ghost fields for the spin $1,\frac{3}{2}$ and $2$ fields.  We list in the table the coefficients $d^{\rm s},c^{\rm s},$ and $f^{\rm s}$, taking into account the contributions of the various ghost fields.\footnote{ Note that: 1) in the table we have presented the coefficients for Weyl (Majorana) fermion, which we obtained by considering a Dirac fermion and divide the result by half;
2) the coefficients $f^{\rm s}$ for gravitino is different from 2; this happens because the contribution of the ghosts, with mass $2m_{3/2}$, is included. For more details see Appendix B. }\\ 


\begin{table}
\begin{center}
\begin{tabular}{|c|c|c|c|cl}
\hline 
${\rm s}$  & $d^{\rm s}$&$c^{\rm s}$&$f^{\rm s}$\tabularnewline
\hline 
\hline 
0  & 29/12&1&1\tabularnewline
\hline 
1/2 & 11/24&-1&2 \tabularnewline
\hline 
1 & -31/6 &-4&2\tabularnewline
\hline 
3/2& 251/24& 8&-88\tabularnewline
\hline 
2  & 1139/6&-22&2\tabularnewline
\hline 
\end{tabular}
\caption{Coefficients appearing in \eqref{CoefflnEm2Gen} for spin $s$ particles.}
\end{center}
\end{table}





\section{Coefficient of \boldmath $\ln |W_0|^2$ in type IIB flux compactifications}\label{sec:logW0}

\subsection{KKLT vacua}

As we have seen, in the KKLT scenario the K\"ahler moduli are fixed by non-perturbative contribution to the superpotential. In this section we consider a Calabi-Yau with one K\"ahler modulus (i.e. $h^{1,1}=1$). The volume $\mathcal{V}$ of the CY will be given in terms of the K\"ahler modulus $\tau$ by $\mathcal{V}=\tau^{3/2}$. We assume that there is a four-cycle $D$ with volume $\tau$ that supports a non-perturbative effect, generating a superpotential of the form $W_{\rm np}=Ae^{-a T}$. 

The $N=1$ supergravity potential is determined by the K\"ahler potential $K$ and the superpotential $W$ of the effective theory. These are functions of the K\"ahler coordinate $T=\tau+i \vartheta$, where $\tau=\frac12 \int_D J^2$ is the K\"ahler modulus and $\vartheta = \int_D C_4$ is the axion coming from the RR four-form potential. After integrating out the complex structure moduli and the axiodilaton, the scalar potential is
\be
V=e^K\l(K^{T\bar T}D_T W\bar{D_T W}-3|W|^2\r) \:.
\ee
In the KKLT case, we have
\be\label{KWKKLT}
K=-2\ln \mathcal{V}(T,\bar T)=-3\ln\l(T+\bar T\r),\qquad W=W_0+Ae^{-a T} \:.
\ee
The supersymmetric minimum of this potential is at $D_T W=0$, i.e. at $\vartheta = 0$ and
\be\label{KKLTmintau}
W_0=-Ae^{-a\tau}\l(1+\tfrac23 a \, \tau \r) \:.
\ee
The value of the potential at the minimum is 
\ben
V|_{\rm min}&=& -\frac{3 W_0^2a^2}{2 \tau \l(3+2a\,\tau\r)^2}
\een
where $\tau$ satisfies the relation \eqref{KKLTmintau}. From this we read the  cosmological constant, i.e. $ \Lambda=V|_{\rm min}$.

\subsubsection*{Scalar masses}

At the minimum, the Hessian of the potential is 
\ben
\partial_i\partial_j V|_{\rm min} &=&\begin{pmatrix} \frac{3W_0^2a^3}{2\tau^2(3+2 a\tau)}&0\\0&\frac{3 W_0^2 a^2 (2+a\tau)(1+2a\tau)}{2\tau^3(3+2a\tau)^2}\end{pmatrix} \:,
\een
with $i,j=\vartheta,\tau$.

We need to calculate the masses of the canonically normalised fields. These are obtained by multiplying the matrix $\partial^2 V$ by $\frac12 K_{T\bar T}^{-1}=\frac{2\tau^2}{3}$. The masses of the two scalar fields are then
\begin{eqnarray}
m_\vartheta^2 &=& \frac{ W_0^2 a^3}{3+2 a \tau} \:, \nn \\
m_\tau^2 &=& \frac{ W_0^2 a^2 (2+a\tau)(1+2a\tau)}{\tau (3+2a\tau)^2} \:.
\end{eqnarray}

\subsubsection*{Fermion mass}
In $N=1$ four dimensional supergravity, the mass matrix for fermion is given by
\ben
m_{ij}^f=m_{3/2}\l(\nabla_iG_j+\frac{1}{3}G_iG_j\r),\qquad G=K+\ln W+\ln\bar W
\een
where $m_{3/2}=e^{K/2}|W|$ is the gravitino mass and
\be
\nabla_i G_j=\p_iG_j-\Gamma^k_{ij}G_k \:,
\ee
with $\Gamma^k_{ij}$ given in \eqref{Gammaijk}.
In the case under study, $i=T$. Moreoever, since we have a susy vacuum, $D_iW=0$. Therefore the fermion mass  is
\be
m^f=m_{3/2} \l[\frac{W_{TT}}{W}+K_{TT}-K_TK_T\r] \:.
\ee
Using \eqref{KWKKLT}, we get
\be
m^f=- \frac{3 W_0 a (1 + a \tau )}{2\sqrt{2} \tau^{5/2}(3+2a\tau)}\:.
\ee
The canonically normalised  mass is
\be
m_{\psi}=- \frac{\sqrt{2} W_0 a (1 + a \tau )}{\tau^{1/2}(3+2a\tau)}\:.
\ee

\subsubsection*{Contribution to $\ln |W_0|^2$}

Now we can calculate the contribution to the logarithmic corrections due to K\"ahler moduli. The contribution due to two scalar fields is
\ben
\Gamma_{(1)}^s=\Big[ \frac{149}{180}+3a\tau  + \frac{25 a^2\tau^2}{6} + \frac{8 a^3\tau^3}{3} + \frac{2 a^4\tau^4}{3}\Big]\ln |W_0|^2 \:.
\een
The corresponding contribution of the fermion is
\ben
\Gamma_{(1)}^f=\Big[\frac{251}{720}+2a\tau + \frac{11 a^2\tau^2}{3} + \frac{8 a^3\tau^3}{3}+\frac{2 a^4\tau^4}{3}\Big]\ln |W_0|^2 \:.
\een
Putting the two results together, the contribution due to a single K\"ahler multiplet is
\be\label{sflnW0KKL}
\Gamma_{(1)}^s-\Gamma_{(1)}^f= \l(-\frac{1}{48}+\frac{1}{2}(1+a\tau)^2\r) \, \ln |W_0|^2=\l(-\frac{1}{48}+\frac{1}{8}m_\psi^2 R_{AdS}^2 \r) \, \ln |W_0|^2  \:,
\ee
where we remind that $R_{AdS}^2=3L^2=\frac{3}{\Lambda}$.
This is the result one expects for a supersymmetric AdS$_4$ minimum, where the scalar masses $m_{s1,s2}$ are determined in terms of the fermion mass\footnote{We refer here to the fermion mass in the canonically normalised Lagrangian.} $m_\psi$ and the radius of AdS $R_{AdS}$ \cite{deWit:1999ui}:
\begin{equation}\label{s1s2fLmasses}
m_{s1,s2}^2 = m_\psi^2-\frac{2}{R_{AdS}^2} \pm \frac{m_\psi}{R_{AdS}}\:.
\end{equation}
If one plug these expressions in \eqref{CoefflnEm2Gen}, the resulting contribution to $\ln |W_0|^2$ matches with \eqref{sflnW0KKL}.
One can also verify that \eqref{s1s2fLmasses} is fulfilled in the present example. 

The contribution coming from the gravity multiplet is a constant, due to supersymmetry. The cosmological constant effectively acts as the mass of the graviton, 
$M_{(2)}^{2}=-2\Lambda=2L^{-2}$, 
while $m_{3/2}=e^{K/2}|W|$ is the mass of the gravitino that in the supersymmetric case is $M_{\{3/2)}^{2}=\frac{1}{3L^2}$ . In this case the contribution to $\ln |W_0|^2$ is given by
\be
\Gamma_{(1)}^m-\Gamma_{(1)}^g=-\frac{113}{48}\ln |W_0|^2
\ee
Notice that in the above derivation, the $e^{K_S+K_{cs}}$ factor in the mass cancels the similar contribution present in $R_{AdS}$.

Summing up all contributions, we obtain
\be\label{Gamma1KKLTOneKm}
\Gamma_{(1)}^{W_0}=\frac{1}{8}\l(-19+ m_\psi^2 R_{AdS}^2\r)\ln |W_0|^2\:.
\ee
Notice that $m_\psi^2 R_{AdS}^2$ is the combination that appears in the relations between masses and conformal dimensions of the dual operators. For the fermion fields, we have $R_{AdS}m_\psi = \Delta_\psi-\frac{d}{2}$ \cite{adscft}.  Hence, in the dual CFT$_3$ the result \eqref{Gamma1KKLTOneKm} can also be written as 
$$\Gamma_{(1)}^{W_0}=\frac{1}{8}\l(-19+ \l(\Delta_\psi-\frac32\r)^2\r)\ln |W_0|^2 \:.$$

Due to the fact the KKLT is supersymmetric, we can immediately write the contribution to $\ln |W_0|^2$ in the case that the Calabi-Yau three-fold $X_3$ has $h^{1,1}$ K\"ahler moduli. Each chiral multiplet associated to a K\"ahler modulus will have a mass scaling like $W_0$ and will give a contribution to $\ln |W_0|^2$ equal to \eqref{sflnW0KKL}. Hence the final result is
\be\label{KKLTGammaW01}
\Gamma_{(1),\,h^{1,1}\,{\rm K. md}}^{W_0}=\l(-\frac{113+h^{1,1}}{48}+\frac{R_{AdS}^2}{8}\sum_{i=1}^{h^{1,1}} m^2_{\psi i} \r)\ln |W_0|^2\:.
\ee


\subsection{LVS vacua}\label{SecLVSStab}

We consider type IIB compactified on a Calabi-Yau (CY) three-fold $X_3$ and take the simplest LVS example, i.e. we take $X_3$ to have two K\"ahler moduli $\tau_b$ and $\tau_s$ and a volume form of swiss cheese type:
\begin{equation}\label{SwChVol}
 \mathcal{V} = \tau_b^{3/2} - \tau_s^{3/2} \:.
\end{equation}

Again the flux superpotetial $W_{flux}$ is generated by switching on three-form fluxes $G_3$. This fixes the complex structure moduli and the axiodilaton at high energies, leaving a constant superpotential $W_0$ at lower energies (depending on the flux numbers). We also assume that the divisor $D_s$ with volume $\tau_s$ supports a non-perturbative effect (like an E3-insanton or a D7-brane stack with a condensing gauge group) generating a  contribution to the superpotential like in KKLT. The total superpotential is then
\begin{equation}
 W = W_0 + A_s e^{-a_s T_s} \:.
\end{equation}
Here $T_s=\tau_s + i \vartheta_s$ is one of the K\"ahler variables of type IIB orientifold compactifications ($T_i= \int_{D_i} (J\wedge J + i C_4)$, with $C_4$ the RR four-form potential).

After integrating out the complex structure moduli and the dilaton, the remaining moduli are the deformations of the K\"ahler form. Their K\"ahler potential (including the leading $\alpha'$-corrections) is
\begin{equation}
K(T_s,T_b) = -2 \log \left(  {\cal V}(T_s,T_b) +  \frac{\xi}{g_s^{3/2}}  \right) \:,
\end{equation}
where $\xi=-\frac{\zeta(3)\chi(X_3)}{4 (2\pi)^3}$.

The scalar potential for the K\"ahler moduli $T_s=\tau_s+i\vartheta_s$ and $T_b=\tau_b+i\vartheta_b$ has a minimum where the volume of $X_3$ is stabilised to be exponentially large. In particular, in the region where ${\cal V}\gg 1$ (i.e. $\tau_b \gg \tau_s$) the potential has the form (after minimizing with respect to the axion $\rho_s$ and taking $W_0 \in \mathbb{R}^{+}$ without loss of generality)
\begin{equation}\label{LVSpot}
 V = \frac{8\,A_s^2a_s^2\sqrt{\tau_s}e^{-2a_s\tau_s}}{3\tau_b^{3/2}} +\cos(a_s \vartheta_s) \frac{4\,A_sa_s W_0\tau_s e^{-a_s\tau_s}}{\tau_b^3} + \frac{3\,W_0^2\xi}{2\,g_s^{3/2} \tau_b^{9/2}} \:.
\end{equation}
 
We see that at this level of approximation, the axion $\vartheta_b$ is a flat direction of the potential. Minimising the potential \eqref{LVSpot} with respect to $\vartheta_s$, $\tau_s$ and $\tau_b$, one obtain the two equations:
\begin{eqnarray}
\label{minEq0}\partial_{\vartheta_s}V=0 &\Leftrightarrow & \vartheta_s= \frac{\pi}{a_s} \\
\label{minEq1}\partial_{\tau_s}V=0 &\Leftrightarrow & \tau_b^{3/2}= \frac{3 e^{a_s\tau_s}W_0\sqrt{\tau_s}(a_s\tau_s -1)}{A_sa_s(4a_s\tau_s-1)} \\
\label{minEq2}\partial_{\tau_b}V=0 &\Leftrightarrow & \frac{g_s^{3/2}}{\xi} = \frac{(4a_s\tau_s-1)^2}{16 a_s\tau_s^{5/2}(a_s\tau_s-1)}
\end{eqnarray}
By restricting to the region in the moduli space where we can trust the supergravity approximation, i.e. $\tau_s$ large, the two minimising equations \eqref{minEq1} and \eqref{minEq2} have the approximated solutions:
\begin{equation}\label{MinConstr}
  {\cal V}\sim \frac{3e^{a_s\tau_s}\sqrt{\tau_s}W_0}{4A_sa_s} \qquad \mbox{and} \qquad \tau_s \sim \frac{\xi^{2/3}}{g_s}
\end{equation}
We see that the volume is stabilized at exponentially large values, as required by the approximation we took at the beginning of the computations. Remember that we are keeping only the leading terms in $1/\tau_b$ expansion. This will hold in the following as well.

By using \eqref{minEq0}, \eqref{minEq1} and \eqref{minEq2}, we can compute the value of the potential at the minimum:
\begin{equation}
 V|_{\rm min} = -\frac{12\, W_0^2 \tau_s^{3/2}(a_s\tau_s-1)}{\tau_b^{9/2}(4\,a_s\tau_s-1)^2}\:.
\end{equation}

\subsubsection*{Scalar masses}

We are now ready to compute the masses of the four real scalar fields $\tau_s,\tau_b,\vartheta_s,\vartheta_b$. 
The masses of the fields are derived by the matrix $\partial_i\partial_j V|_{\rm min}$. In our case this matrix is block-diagonal. The block relative to the axions $\vartheta_b,\vartheta_s$ is (at leading order in the $1/\tau_b$ expansion)
\begin{eqnarray}
  \partial_{\vartheta_{\bar{j}}}\partial_{\vartheta_k}V|_{\rm min}&=& \tfrac{6W_0^2(a_s\tau_s-1)}{\tau_b^{9/2}(4a_s\tau_s-1)} \left(
 \begin{array}{cc}
0 & 0 \\
0 & 2a_s^2\tau_s^{3/2}\\
\end{array}\right) \:. 
\end{eqnarray}
while the block relative to $\tau_b,\tau_s$ is
\begin{eqnarray}
  \partial_{\tau_{\bar{j}}}\partial_{\tau_k}V|_{\rm min}&=& \tfrac{6W_0^2(a_s\tau_s-1)}{\tau_b^{9/2}(4a_s\tau_s-1)} \left(
 \begin{array}{cc}
\frac{9\tau_s^{3/2}(2a_s\tau_s + 1)}{\tau_b^2(4a_s\tau_s-1)} & -\frac{3\tau_s^{1/2}(a_s\tau_s-1)}{\tau_b} \\
-\frac{3\tau_s^{1/2}(a_a\tau_s-1)}{\tau_b} & \frac{1+3a_s\tau_s-6a_s^2\tau_s^2+8a_s^3\tau_s^3}{\tau_s^{1/2}(4a_s\tau_s-1)} \\
\end{array}\right) \:. 
\end{eqnarray}

We are interested in the canonically normalised fields. The masses are the eigenvalues of the matrices $\frac12 K^{i\bar{j}}\partial_{\vartheta_{\bar{j}}}\partial_{\vartheta_k}V|_{\rm min}$ and $\frac12 K^{i\bar{j}}\partial_{\tau_{\bar{j}}}\partial_{\tau_k}V|_{\rm min}$.
The inverse of the K\"ahler metric is (at leading order in the $1/\tau_b$ expansion)
\begin{eqnarray}
 K^{i\bar{j}}|_{\rm min} &=& \left(
 \begin{array}{cc}
 \frac{4}{3}\tau_b^2 & 4 \tau_b\tau_s \\ 4 \tau_b\tau_s & \frac{8}{3}\tau_b^{3/2}\tau_s^{1/2}
 \end{array}\right)    
\end{eqnarray}

The eigenvalues of the matrix $K^{i\bar{j}}\partial_{\bar{j}}\partial_kV|_{\rm min}$ gives the physical masses of the canonically normalised fields:
\begin{eqnarray} 
 m^2_{\Theta} &=& 0 \\ \nonumber \\
 m^2_{\theta} &=&  \frac{16W_0^2 a_s^2 \tau_s^2 (a_s\tau_s-1)}{\tau_b^3 (4a_s\tau_s-1)}  \\ \nonumber \\
 m^2_{\Phi} &=& \frac{108 W_0^2 a_s \tau_s^{5/2}(a_s\tau_s-1)(5-11a_s\tau_s+12a_s^2\tau_s^2) }{\tau_b^{9/2}(4a_s\tau_s - 1)^2(1+3a_s\tau_s-6a_s^2\tau_s^2+8a_s^3\tau_s^3)} \\ \nonumber \\
 m^2_{\phi} &=& \frac{8 W_0^2(a_s\tau_s - 1)(1+3a_s\tau_s-6a_s^2\tau_s^2+8a_s^3\tau_s^3)}{\tau_b^3 (4 a_s\tau_s-1)^2}
\end{eqnarray}
We immediately realise that $m^2_{\phi},m^2_{\theta}\gg \frac{1}{L^2}$, while $m^2_{\Phi} $ is of the same order as $\frac{1}{L^2}$.

We can approximate the values of $\frac{1}{L^2}$, $m^2_{\Phi}$ and $m^2_{\phi}$ in the limit $a_s\tau_s \gg 1$. This is a valid approximation. In fact $a_s\sim 1$, while $\tau_s\sim \frac{\xi^{2/3}}{g_s}$: to be in a controlled regime $g_s\ll 1$ (in the explicit example presented below,  $\xi\sim 2.08$). In this approximation

\begin{eqnarray}
\frac{1}{L^2}   &=&   \frac{3 W_0^2\tau_s^{1/2}}{4 \tau_b^{9/2} a_s} \left(1 + \frac{1}{2a_s\tau_s} + ...         \right) \\
m^2_{\theta}   &=&  \frac{4W_0^2 a_s^2 \tau_s^{2}}{\tau_b^3} \left(1 - \frac{3}{4a_s\tau_s} + ...          \right) \\
m^2_{\Phi}   &=&   \frac{81W_0^2\tau_s^{1/2}}{8 \tau_b^{9/2} a_s} \left(1 - \frac{2}{3a_s\tau_s} + ...         \right) \\
m^2_{\phi}   &=&   \frac{4W_0^2a_s^2\tau_s^2}{\tau_b^3} \left(1 - \frac{5}{4a_s\tau_s} + ...         \right) 
\end{eqnarray}
We see that at leading order in this approximation, we have $m^2_{\Phi}=\frac{27}{2L^2}$ and $m_\phi = m_\theta$.

\subsubsection*{Fermion masses}

Let us now compute the masses for the (canonically normalised) modulini.
We start from the fermion mass matrix in the sugra sigma model:
\ben
m^f_{ij}=m_{3/2}\l(\nabla_iG_j+\frac{1}{3}G_iG_j\r),\qquad G=K+\ln W+\ln\bar W
\een
where $m_{3/2}=e^{K/2}|W|$ is the gravitino mass and $\nabla_i G_j=\p_iG_j-\Gamma^k_{ij}G_k$.
For the present case, this matrix reads
\begin{equation}
 m^f_{ij} \,=\, -\frac{3W_0}{8\tau_b^3}\left(
 \begin{array}{cc}
 \frac{\tau_s^{3/2}(2 a_s \tau_s+7)}{\tau_b^2(4a_s\tau_s-1)^2} & - \frac{3\tau_s^{1/2}(2a_s\tau_s-1)}{\tau_b (4a_s \tau_s-1)} \\ - \frac{3\tau_s^{1/2}(2a_s\tau_s-1)}{\tau_b (4a_s \tau_s -1)} & \frac{(2a_s \tau_s -1)}{\tau_s^{1/2}}
 \end{array}\right)    
\end{equation}
As for the scalars, we compute the canonically normalised masses for the two mass eigenstates:
\begin{eqnarray}
m_{\Psi}   &=&   - \frac{W_0 (2a_s\tau_s -1)}{\tau_b^{3/2}} = - \frac{2 W_0 a_s\tau_s }{\tau_b^{3/2}} \left(1 - \frac{1}{2a_s\tau_s} + ...\right)  \\
m_{\psi}   &=&    \frac{8 W_0 \tau_s^{3/2}(a_s\tau_s -1)}{\tau_b^3 (4 a_s\tau_s-1)^2} = \frac{ W_0 \tau_s^{1/2}}{2a_s\tau_b^3}\left( 1-\frac{1}{2a_s\tau_s} + ... \right)
\end{eqnarray}

\subsubsection*{Contribution to $\ln |W_0|^2$}

We can now compute the contribution to $\ln |W_0|^2$ coming from the K\"ahler moduli spectrum. Like in KKLT, we assume that there are no further massless fields remaining.

The scalar contribution coming from the four scalars is at leading order in the $1/\tau_b$ expansion: 
\begin{eqnarray}
\Gamma^s_{(1)} &=& \frac{\tau_b^3(1+6 a_s \tau_s-3a_s^2\tau_s^2-20 a_s^3\tau_s^3+88a_s^4\tau_s^4-128a_s^5\tau_s^5+128a_s^6\tau_s^6)}{12\tau_s^3} \ln |W_0|^2
\end{eqnarray}
The leading contribution in the $\tau_b$ expansion is basically given by the $m^4$ term relative to the fields $\theta$ and $\phi$. In fact, their masses scales with powers of $\tau_b$ with respect to the $1/L$, i.e. $L\cdot m_{\theta,\phi} \sim \tau_b^{3/4}$.
 
The fermion contribution is basically given at leading order in $\tau_b$ by the $m_\Psi^4$ term:
\begin{eqnarray}
\Gamma^f_{(1)} &=& \frac{\tau_b^3 (4 a_s \tau_s -1)^4(2a_s \tau_s -1)^4}{384\tau_s^3(a_s\tau_s-1)^2} \ln |W_0|^2 \:.
\end{eqnarray}

Considering both contribution, we obtain
\begin{equation}\label{CoeffW0LVSs-f}
\Gamma^s_{(1)} - \Gamma^f_{(1)} =  \frac{\tau_b^3(31+152 a_s \tau_s - 696 a_s^2\tau_s^2 + 1184 a_s^3\tau_s^3 - 1136 a_s^4\tau_s^4 + 1152 a_s^5\tau_s^5 - 768 a_s^6\tau_s^6)}{384\tau_s^3(a_s\tau_s-1)^2}\ln |W_0|^2 \:.
\end{equation}

The gravity multiplet contributes differently with respect to the KKLT.
Since the minimum is not supersymmetric, the gravitino contribution is not determined by the graviton one. 
In this case the contribution to $\ln |W_0|^2$ is given by
\be
\Gamma_{(1)}^m-\Gamma_{(1)}^g= \frac{11\tau_b^3(4a_s\tau_s-1)^4}{96\tau_s^3(a_s\tau_s-1)^2}\ln |W_0|^2
\ee
The $\tau_b$ dependence comes from the gravitino mass, whose $\tau_b$ scaling is different from the one of $1/L$. This is a difference with respect to what happens in the KKLT case.

If we sum up all the contribution, we obtain
\begin{equation}\label{LVSGammaW01}
 \Gamma^{W_0}_{(1)} = \frac{\tau_b^3(25-184 a_s \tau_s +1176 a_s^2\tau_s^2 -3360 a_s^3\tau_s^3 + 3376a_s^4\tau_s^4 + 384 a_s^5\tau_s^5 - 256 a_s^6\tau_s^6)}{128\tau_s^3(a_s\tau_s-1)^2}\ln |W_0|^2 \:.
\end{equation}
Taking the leading term in the $\tau_s\gg 1$ limit, we obtain 
\begin{equation}\label{LVSGammaW02}
 \Gamma^{W_0}_{(1)} \sim - 2\, a_s^4 \, \tau_b^3 \, \tau_s \, \ln |W_0|^2 \:.
\end{equation}
This leading contribution comes from $\Gamma_{(1)}^s-\Gamma_{(1)}^f$, as the gravity contribution is subleading for $\tau_s\gg 1$.

\subsubsection*{A simple global model}

We present an explicit global model for a LVS minimum, i.e. we consider an explicit Calabi-Yau threefold and an orientifold projection, with a setup of branes that satisfies all the string theory consistency conditions (like tadpole cancellation, proper quantisation of fluxes, etc...).
The compactification manifold $X_3$ is the famous CY $\mathbb{P}^4_{11169}[18]$. More precisely, it is an hypersurface  described by the vanishing locus of a polynomial of degrees $(18,6)$ in the toric ambient variety defined by the following weights 
\begin{equation}
\begin{array}{cccccc}
u_1 & u_2 & u_3 & x & y & z\\ 
\hline 
1 & 1 & 1 & 6 & 9 & 0\\
0 & 0 & 0 & 2 & 3 & 1\\
\end{array}
\end{equation}
and with SR-ideal given by $\{\,u_1u_2u_3,x\,y\,z\,\}$.
This Calabi-Yau manifold has Hodge numbers $h^{1,1}=2$ and $h^{1,2}=272$, with Euler characteristic $\chi(X_3)=-540$.

An integral basis of divisor is given by $D_1,D_z$ (with $D_1=\{u_1=0\}$ and $D_z=\{z=0\}$), with intersection numbers
\begin{equation}
D_1^3=0 \qquad D_1^2 D_z=1 \qquad D_1D_z^2=-3 \qquad D_z^3=9 \:.
\end{equation}
We expand the K\"ahler form in the basis of Poincar\'e dual two forms $\hat{D}_1,\hat{D}_z$: $J=t_1 \hat{D}_1 + t_z\hat{D}_z$.
The volumes of the divisors $D_z$ and $D_y=9D_1+3D_z$ are
\begin{equation}
 \tau_z=\tfrac12\int_{D_z}J^2=\tfrac12(t_1-3t_z)^2 \qquad  \tau_y=\tfrac12\int_{D_y}J^2=\tfrac32t_1^2\:,
\end{equation}
while the volume of the CY is
\begin{equation}
 {\cal V} = \frac16\int_{X_3}J^3 = \frac{1}{18} \left( t_1^3-(t_1-3t_z)^3 \right) = \frac{\sqrt{2}}{9}\left( \left(\frac{\tau_y}{3}\right)^{3/2}-\tau_z^{3/2} \right)\:.
\end{equation}
In the following we will use the variables $\tau_b\equiv \tau_y/3$ and $\tau_s\equiv \tau_z$. Thevolume of $X_3$ takes then the form
\begin{equation}
 {\cal V} = \frac{\sqrt{2}}{9}\left( \tau_b^{3/2}-\tau_s^{3/2} \right)\:.
\end{equation}
We note that this is equal to \eqref{SwChVol}, up to the overall factor. This can be absorbed into a rescaling of $W_0,A_s,\xi$. In detail, this model is equivalent to the one described by the volume form \eqref{SwChVol}, if
$W_0\mapsto \frac{9}{\sqrt{2}}W_0 $ and $A_s\mapsto \frac{9}{\sqrt{2}}A_s $ and the definition of $\xi$ is also rescaled $\xi \mapsto \frac{9}{2\sqrt{2}}\xi$. The new $\xi$ is equal to $\xi\sim 2.08$ in this model (where we have used $\chi(X_3)=-540$).

The only other (non-flux dependent) parameter in the scalar potential that remains to be determined is $a_s$. It depends on the non-perturbative effects that lives on the four-cycle $D_s=D_z$.
We consider two cases, corresponding to two different orientifold involutions. These lead to a different spectrum and different nature of the non-perturbative effect. 
\begin{itemize}
\item[$1)$] The orientifold involution is given by
\begin{equation}
\sigma\,: \qquad  z \mapsto -z \:.
\end{equation}
The fixed point locus is made up of two O7-planes at $z=0$ and $y=0$. They do not intersect each other. The orientifold-plane D7-tadpole is cancelled by taking four D7-branes (plus their four images) on top of $z=0$ and a fully recombined D7-brane wrapping a 4-cycle in the homology class $8D_y$ (called in litterature 'Whitney brane' for its characteristic shape) \cite{DenefCollinucci}. 
The stack on $z=0$ gives an $SO(8)$ gauge group, while the Whitney brane does not support any gauge symmetry.
We choose a background value for the bulk B-field equal to $B=\frac{\hat{D}_z}{2}$. In this way there is a choice of gauge flux on the D7-branes such that the gauge invariant flux ${\cal F}=F-\iota^\ast B$ can be set to zero. In fact, Freed-Witten anomaly cancellation requires the gauge flux on the branes on $z=0$ to be half-integrally quantized ($F+\frac{c_1(D_z)}{2} \in H^2(D_z,\mathbb{Z})$). 
With techniques described in \cite{Collinucci:2008sq,Cicoli:2011qg} one can compute the D3-charge of this configuration. We make a choice of the flux on the Whitney brane that maximize the absolute value of the charge, obtaining $Q_{D3}^{D7}=1491$. 
This large negative contribution to $Q_{D3}$ allows to switch on positively contributing three-form fluxes on the bulk and two-form fluxes on the Whitney brane; these stabilise at large scale the complex structure moduli, the axiodilaton and the open string moduli describing the deformations of the Whitney brane \cite{DenefCollinucci}.

By using proper index theorems, one can compute (see for example \cite{compact4}) the number of even and odd (1,2)-forms on $X_3$. With the chosen orientifold involution, we have $h^{1,2}_+=0$ and hence $h^{1,2}_-= h^{1,2}=272$. This means that we have no massless gauge multiplet coming from $C_4$ expanded on even three-forms.

The divisor $D_z$ is a rigid $\mathbb{CP}^2$ and hence it has $h^{1,0}=h^{2,0}=0$. This means that the theory living on the corresponding D7-brane stack is a pure $SO(8)$ SYM. It undergoes gaugino condensation, generating a superpotential
\begin{equation}\label{NPsuperpot}
 W_{\rm np} = A_s e^{-a_s T_s} \:.
\end{equation}
with $a_s=\pi/3$.

\item[$2)$] The orientifold involution is given by
\begin{equation}
\sigma\,: \qquad  x \mapsto -x \:.
\end{equation}
The fixed point locus is made up of one O7-plane at $x=0$. The orientifold-plane D7-tadpole is cancelled by a Whitney brane wrapping a four-cycle in the homology class $8D_x$. Hence we do not have any massless guage multiplet coming from the D7-brane worldvolume.
The D3-charge of the D7-brane and the O7-plane (considering zero flux on the D7-brane) is $Q_{D3}^{D7}=498$. 


By using the index theorems, we compute $h^{1,2}_+=69$ and $h^{1,2}_-=203$. This means that we have $n_{\rm gauge}=69$ massless gauge multiplets. These will contribute to the coefficient of $\ln |W_0|^2$ with a constant term that is subleading with respect to the \eqref{CoeffW0LVSs-f}.

The rigid divisor $D_z$ is not wrapped by any D7-brane. On the other hand, an invariant E3-instanton is wrapped on $D_z$ when $B=\frac{\hat{D}_z}{2}$. This will contribute to the non-perturbative superpotential $W_{\rm np}=A_s e^{-a_s\tau_s}$, with $a_s= 2\pi$. If $B=0$, the leading contribution will be given by E3-instantons with higher rank, as described in \cite{InakiPer}.

\end{itemize}

Inserting the model-dependent value of $a_s$ into \eqref{LVSGammaW01} (or \eqref{LVSGammaW02}), one obtains the coefficient of $\ln |W_0|^2$ in terms of $\mathcal{V}$ and $g_s$.\footnote{Unfortunately the lack of knowledge of the explicit expression of the prefactor $A_s$ in terms of the complex structure moduli, does not allow to obtain explicit numbers for the coefficient of $\ln |W_0|^2$ (as we vary the fluxes to follow the limit $W_0\rightarrow 0$). This is still true also in the subset of flux vacua considered in \cite{MartinezPedrera:2012rs,compact4}, where by switching on only symmetric fluxes, the values of $g_s$ and $W_0$ could be computed.}

\subsection*{Summary}

Our main results in this section are given by equations (\ref{KKLTGammaW01}) for KKLT and (\ref{LVSGammaW01}) for LVS. These results are obtained by one loop calculations in the supergravity coupled to K\"ahler moduli about AdS$_4$ background. In these computation we expressed one loop determinant of the differential operator in terms of heat kernel and then considered it's small $\tau$ expansion, see \eqref{eq:GammAdiabatic}. The coefficients of the $\tau$ expansion are expressed in terms of curvature invariants and masses of fields, that we derive explictly for both KKLT and LVS. From such an expansion we extracted the coefficient of $\ln|W_0|$ which is given by the coefficient of $\tau$ independent term in the heat kernel expansion. We computed these contributions in the limit  (\ref{limitW0}) where $|W0|$ is taken to be small while keeping $g_s$ and $\mathcal{V}$ fixed. In this limit we can ignore the contribution of the complex structure moduli and KK modes, whose masses do not scale with $W_0$.\footnote{In other limits, where also $\mathcal{V}$ and $g_s$ vary, we shoul include these masses, that are not computable with present techniques in a generic flux compactification.} In the AdS/CFT dictionary the $\ln |W_0|$ term will correspond to a term $\sim\log c$ in the free energy of the dual CFT in the $\frac{1}{N}$-expansion. Thus our calculation provide a non trivial consistency checks for any candidate CFT dual.

\section{Discussion}

In this paper we have made  some progress in describing the properties of the CFT duals of  AdS vacua of KKLT and LVS type. Our main technical result is the identification of a concrete calculation, that we performed, of a duality independent quantity. This is the coefficient of the logarithmic term of the one-loop vacuum energy. For the KKLT case the result is quite simple 
and depends only on the conformal dimension of the involved K\"ahler moduli and on $h^{1,1}$.  
For the LVS case it is a model dependent quantity depending on the values of the moduli at the minimum. The difference relies on the fact that the KKLT AdS vacua preserve supersymmetry whereas in the LVS case supersymmetry is spontaneously broken. In both cases we present then a concrete prediction that in principle should be computable once a CFT dual candidate is identified. 
Performing the equivalent calculation on the CFT side is left as an outstanding open question since we still have very limited information on the CFT duals. For example, one would need to know, among other features, the parameters (or the combinations of the parameters) of the CFT that corresponds to $W_0$, $g_s$ and $A$ (or $\tau_s$ and $\tau_b$ in LVS). Only after that  can one select the $\ln |W_0|^2$ term in the partition function and check the coefficient.


Our results are a small step towards identifying the CFT duals of the landscape of AdS vacua and therefore towards its proper non-perturbative formulation. They  could also lead to  applications. The three dimensional CFT duals that we have tried to uncover could provide good candidates for some of the applications of AdS/CFT duality. In particular the non-supersymmetric LVS vacua could be relevant for studies of condensed matter applications. The fact that these non-supersymmetric CFTs are particularly simple with only one scalar operator with $\mathcal{O}(1)$ conformal dimension  may give rise to interesting implications.

There are many questions left open. A typical chiral model with moduli stabilised has many ingredients that should have a counterpart on the CFT side. Besides string, Kaluza-Klein and moduli states, chiral visible and hidden sectors are present with a diversity of gauge and matter fields which are model dependent but have to manifest themselves in the dual theory. In general essentially all the compact models have anomalous $U(1)$s with anomaly cancelled by the Green-Schwarz mechanism. These gauge fields get a mass by the Stuckelberg mechanism. It may be interesting to find the dual realisation of this mechanism which is generic in string compactifications. A proper understanding of supersymmetry breaking on the CFT side would also be desirable.
  
Besides the AdS vacua studied here, the string landscape also  includes de Sitter solutions. A typical potential will have minima with both signs of the cosmological constant and transitions between them should be approached from the dual side. These dS solutions are less understood but would be interesting to explore, extending some of the discussions in this article.

\section*{Acknowledgements}
We thank useful conversations with Luis Aparicio, Matteo Bertolini, Michele Cicoli, Joe Conlon, Atish Dabholkar, Justin David, Oliver DeWolfe, Anshuman Maharana, Juan Maldacena, Ashoke Sen, Marco Serone, Kyriakos Papadodimas, Giovanni Villadoro and Matthew Williams.
SdA wishes to thank the Abdus Salam ICTP for hospitality during the initial stages of this project and his research
is partially supported by the United States Department of Energy under
grant DE-FG02-91-ER-40672.

\appendix

\section{\boldmath$\mathcal N=1$ supergravity Lagrangian}

The supergravity Lagrangian in our conventions (MTW) is 
\ben
\mathcal L&=&\frac{1}{2}R-g_{i\bar j}\p_\mu \phi^i\p^\mu\bar\phi^{\bar j}-ig_{i\bar j}\bar\chi^{\bar j}\bar\sigma^\mu \mathcal D_\mu\chi^i+\varepsilon^{klmn}\bar\psi_k\bar\sigma_l\tilde{\mathcal D_m}\psi_n\nn\\&&-\frac{1}{\sqrt{2}}g_{i\bar j}\p_n\bar\phi^{\bar j}\chi^i\sigma^m\bar\sigma^n\psi_m-\frac{1}{\sqrt{2}}g_{i\bar j}\p_n\phi^{i}\bar\chi^{\bar j}\bar\sigma^m\sigma^n\bar\psi_m-e^{G/2}\Big\{\psi_a\sigma^{ab}\psi_b+\bar\psi_a\bar\sigma^{ab}\bar\psi_b\nn\\&&+\frac{i}{\sqrt{2}}G_i\chi^i\sigma^a\bar\psi_a+\frac{i}{\sqrt{2}}\bar G_{\bar i}\bar\chi^{\bar i}\bar\sigma^a\psi_a+\frac{1}{2}[G_{ij}+G_iG_j-\Gamma^k_{ij}G_k]\chi^i\chi^j+\frac{1}{2}[\bar G_{\bar i\bar j}\nn\\&&+\bar G_{\bar i}\bar G_{\bar j}-\bar\Gamma^k_{ij}\bar G_{\bar k}]\bar\chi^i\bar\chi^j\Big\}-e^G[g^{i\bar j}G_i\bar G_j-3]
\een
In the above we have
\be
G=K+\ln W+\ln \bar W
\ee
Also in the above Christoffel connection is defined as
\be\label{Gammaijk}
\p_kg_{i\bar j}=g_{m\bar j}\Gamma^m_{ik}.
\ee 


\section{One loop computation}\label{AppB}
The calculations below are based on the deWitt coefficients given in \cite{Christensen:1978md, Vassilevich:2003xt}. 
\subsection{Scalar field}
For a scalar field we have the Lagrangian \footnote{Note that all calculations are done in a Euclidean metric},
\be
\mathcal{L}_{scalar}=\frac{1}{2}\phi\l[-\Box+m_s^2\r]\phi .
\ee
For a massless scalar field we have the following deWit coefficients
\be
\boldsymbol{a}_0=1,\quad \boldsymbol{a}_2=\frac{1}{6}R,\quad \boldsymbol{a}_4=\frac{1}{180}\l[C_{\mu\nu\rho\sigma}C^{\mu\nu\rho\sigma}+\l(R_{\mu\nu}R^{\mu\nu}-\frac{1}{4}R^2\r)+\frac{29}{12}R^2\r].
\ee
Therefore for a massive scalar field, we have
\ben
\boldsymbol{a}_{4}(total)&=&\frac{1}{2}\boldsymbol{a}_{0}m_s^{4}-\boldsymbol{a}_{2}m_s^{2}+\boldsymbol{a}_{4}\nn \\
&=& \frac{1}{2}m_s^4-\frac{1}{6}m_s^2R+\frac{1}{180}\l[C_{\mu\nu\rho\sigma}C^{\mu\nu\rho\sigma}+\l(R_{\mu\nu}R^{\mu\nu}-\frac{1}{4}R^2\r)+\frac{29}{12}R^2\r].\,
\een
\subsection{Vector field}
Let us first consider a $U(1)$ gauge field with Lagrangian
\be
\mathcal{L}_{vector}=\frac{1}{4}F_{\mu\nu}F^{\mu\nu}.
\ee
We need to add a gauge fixing term
\be
\mathcal{L}_{g.f.}=\frac{1}{2}(\nabla_\mu A^\mu)^2.
\ee
The total Lagrangian is
\be
\mathcal{L}_{vector}+\mathcal{L}_{g.f.}=-\frac{1}{2}A_\mu\l(-\Box g^{\mu\nu}+R^{\mu\nu}\r)A_\nu .
\ee
We also need to include the contribution of two ghost field. Thus the total contribution to deWitt coefficients are given by
\be
\boldsymbol{a}_0=2,\quad \boldsymbol{a}_2=-\frac{4}{6}R,\quad \boldsymbol{a}_4=\frac{1}{180}\l[-13C_{\mu\nu\rho\sigma}C^{\mu\nu\rho\sigma}+62\l(R_{\mu\nu}R^{\mu\nu}-\frac{1}{4}R^2\r)-\frac{31}{6}R^2\r].
\ee
\subsection{Graviton}
We consider the Lagrangian of the form
\be
\mathcal{L}=-\frac{1}{2}\l(R-2\Lambda\r).
\ee
In this section we will follow the calculation presented in \cite{Christensen:1979iy}. Since graviton has gauge degree of freedom, we need to add gauge fixing term and also ghost term in the Lagrangian. We use harmonic gauge in which we 
\be
\nabla^\mu\phi_{\mu\nu}=0,\quad \phi_{\mu\nu}=h_{\mu\nu}-\frac{1}{4}g_{\mu\nu}h^\mu{}_\mu .
\ee
Also the ghost is the grassmann valued vector field $\phi_\mu$ and its Lagrangian is
\be
\mathcal L_{ghost}=\phi_\mu^*(-g^{\mu\nu}\Box-R^{\mu\nu})\phi_\nu .
\ee  
At the quadratic order the complete action is given by
\be
S=-\int d^4x\sqrt{g}\l[\frac{1}{2}\phi^{\mu\nu}\Delta^\Lambda(1,1)\phi_{\mu\nu}-\frac{1}{2}\phi\Delta^\Lambda(0,0)\phi+\phi_\mu^*\Delta^\Lambda(\frac{1}{2},\frac{1}{2})\phi^\mu\r] ,
\ee
where
\ben
&&\Delta^\Lambda(1,1)\phi_{\mu\nu}=-\nabla^\rho\nabla_\rho\phi_{\mu\nu}-2R_{\mu\rho\nu\sigma}\phi^{\rho\sigma}\nn \\
&&\Delta^\Lambda(\frac{1}{2},\frac{1}{2})\phi_{\mu}=-\nabla^\rho\nabla_\rho\phi_{\mu}-\Lambda\phi^{\mu}\\
&&\Delta^\Lambda(0,0)\phi=-\nabla^\rho\nabla_\rho\phi-2\Lambda\phi .\nn
\een
In the above $\phi_{\mu\nu}$ is the traceless part of $h_{\mu\nu}$ and $\phi$ is the trace part. Thus including the contribution of ghost field, we get the following deWitt coefficients 
\ben
&&\boldsymbol{a}_{0}=2,\quad \boldsymbol{a}_{2}=-\frac{22}{6}R,\nn\\&&\boldsymbol{a}_{4}=\frac{1}{180}\l(212C_{\mu\nu\rho\sigma}C^{\mu\nu\rho\sigma}+\frac{1139}{6}R^2\r)
\een
Since the cosmological constant effectively acts as the mass for the graviton, the total $\boldsymbol{a}_{4}$ including the contribution of the effective mass is given by
\be
\boldsymbol{a}^\Lambda_{4}(total)=\boldsymbol{a}_{4}+2\Lambda\boldsymbol{a}_{2}+2\Lambda^2\boldsymbol{a}_{0}
\ee
\subsection{Dirac fermion}
The fermonic Lagrangian is
\be
\mathcal L_{fermion}=-i\bar\psi\bar\sigma^\mu D_\mu\psi-\frac{1}{2}m\psi\psi-\frac{1}{2}m\bar\psi\bar\psi
\ee
In the above action $\psi$ is a chiral fermion, $\bar\sigma_\mu=(\mathds I,-\vec \sigma)$, $\vec \sigma$ are Pauli matrices. Now the above can be further written as
\ben
\mathcal L_{fermion}&=&-\frac{i}{2}\bar\psi\bar\sigma^\mu D_\mu\psi-\frac{i}{2}\psi\sigma^\mu D_\mu\bar\psi-\frac{1}{2}m\psi\psi-\frac{1}{2}m\bar\psi\bar\psi\nn\\&&=-\frac{1}{2}\bar\Psi(i\Gamma^\mu D_\mu+m)\Psi
\een
In the above
\be
\Psi=\begin{pmatrix}\psi\\\bar\psi\end{pmatrix},\quad \Gamma_\mu=\begin{pmatrix}0&\sigma_\mu\\\bar\sigma_\mu&0\end{pmatrix},\quad D_\mu\Psi=\p_\mu\Psi+\frac{1}{8}\omega_\mu^{ab}[\Gamma_a,\Gamma_b]\Psi
\ee
Here $\Gamma$ matrices satisfy the Clifford algebra
\be
\{\Gamma^a,\Gamma^b\}=-2\eta^{ab},\quad \eta_{ab}=(-1,+1,+1,+1)
\ee
The above gamma matrix satisfy
\be
\Gamma^{a\dagger}\Gamma^0=\Gamma^0\Gamma^a
\ee
Defining the gamma matrix $\gamma^\mu$ as
\be
\gamma^a=i\Gamma^a,\quad \{\gamma^a,\gamma^b\}=2\eta^{ab},\quad \gamma^{a\dagger}\gamma^0=-\gamma^0\gamma^a
\ee

We can rewrite the above Lagrangian as
\be
\mathcal L_{fermion}=-\frac{1}{2}\bar\Psi(\gamma^\mu D_\mu+m)\Psi
\ee
Now we do analytic continuation to Euclidean space. In this case we assume that $\bar\psi$ is indep. of $\psi$ and hence $\Psi$ is a Dirac spinor. We calculate the one loop determinant and divide the result by half as we are doubling the number of degrees of freedom.\\
We note that in Euclidean space $\gamma^{\mu\dagger}=\gamma^\mu$.
 Then the one loop determinant is
\ben
\ln\mathcal Z_{fermn}&\sim& \ln det(\gamma^\mu D_\mu+m)\sim \frac{1}{2}\ln det(\gamma^\mu D_\mu+m)det(-\gamma^\mu D_\mu+m)\nn\\&&\sim\frac{1}{2}\ln det(-\Box\mathds{1}+m^2-\gamma^\mu\gamma^\nu D_{[\mu} D_{\nu]})
\een
Now
\be
\gamma^\mu\gamma^\nu D_{[\mu} D_{\nu]}\psi=-\frac{1}{4}R\psi,\quad D_{[\mu} D_{\nu]}=\frac{1}{2}\l[D_\mu D_\nu-D_\nu D_\mu\r]
\ee
Here $R$ is the Ricci scalar. 
Thus for the massless Dirac fermion, we get
\ben
&&\boldsymbol{a}_{0}=4,\quad \boldsymbol{a}_{2}=-\frac{2}{6}R, \nn\\&&\boldsymbol{a}_{4}=\frac{2}{180}\l[\frac{11}{24}R^2-\frac{11}{2}\l(R_{\mu\nu}R^{\mu\nu}-\frac{1}{4}R^2\r)-\frac{7}{4}C_{\mu\nu\rho\sigma}C^{\mu\nu\rho\sigma}\r]
\een
In this case for massive Dirac fermion we have
\ben
\boldsymbol{a}_{4}(total)&=&\frac{1}{2}\boldsymbol{a}_{0}m^4-\boldsymbol{a}_{2}m^2+\boldsymbol{a}_{4}\nn\\&=&\frac{1}{90}\l[180\l(m^2+\frac{R}{4}\r)\l(m^2-\frac{R}{12}\r)+\frac{101}{24}R^2-\frac{11}{2}\l(R_{\mu\nu}R^{\mu\nu}-\frac{1}{4}R^2\r)-\frac{7}{4}C_{\mu\nu\rho\sigma}C^{\mu\nu\rho\sigma}\r]\nn\\
\een
Since we have computed the determinant for Dirac fermion, in order to get the result for Weyl/Majorana fermion we have to divide the above result by half. 
\subsection{Gravitino }
Next we consider the Lagrangian for gravitino
\be
\mathcal L_{gravitino}=\epsilon^{klmn}\bar\psi_k\bar\sigma_l\tilde{\mathcal D}_m\psi_n-m_\psi[\psi_a\sigma^{ab}\psi_b+\bar\psi_a\bar\sigma^{ab}\bar\psi_b]
\ee
Here
\be
\tilde{\mathcal D}_m\psi_n=\p_m\psi_n+\frac{1}{8}\omega_m^{ab}[\Gamma_a,\Gamma_b]\psi_n-\tilde\Gamma^k_{mn}\psi_k+\frac{1}{4}(K_j\p_m\phi^j-K_{\bar j}\p_m\bar \phi^j)\psi_n
\ee
For our background, the last term is zero as the scalar fields are constant. The above Lagrangian can also be written as
\be
\mathcal L_{gravitino}=\frac{1}{2}\epsilon^{klmn}\bar\psi_k\bar\sigma_l\tilde{\mathcal D}_m\psi_n+\frac{1}{2}\epsilon^{klmn}\psi_n\sigma_l\tilde{\mathcal D}_m\bar\psi_k-m_\psi[\psi_a\sigma^{ab}\psi_b+\bar\psi_a\bar\sigma^{ab}\bar\psi_b]
\ee
Now we define a Dirac spinor and $\Gamma^{\mu\nu}$ as
\be
\Psi_m=\begin{pmatrix}\psi_m\\\bar\psi_m\end{pmatrix},\qquad \Gamma^{\mu\nu}=\frac{1}{2}\l[\Gamma^\mu,\Gamma^\nu\r]
\ee
Then the above Lagrangian can be written as
\be
\mathcal L_{gravitino}=\frac{1}{2}\epsilon^{klmn}\bar\Psi_k\Gamma_l\Gamma_5\tilde{\mathcal D}_m\Psi_n+\frac{1}{2}m_\psi\bar\Psi_a\Gamma^{ab}\Psi_b
\ee
In the above 
\be
\Gamma_5=i\Gamma_0\Gamma_1\Gamma_2\Gamma_3=\begin{pmatrix}{\mathds 1_{2\times 2}}&0\\0&-{\mathds 1_{2\times 2}}\\\end{pmatrix}
\ee
The gauge transformation can be written as
\be
\delta\Psi_\mu=2D_\mu\hat\epsilon+im_\psi\Gamma_\mu\hat\epsilon,\quad \hat\epsilon=\begin{pmatrix}\epsilon\\\bar\epsilon\end{pmatrix}
\ee
Using the relation
\be
\Gamma^{\mu\nu\rho}=-i\epsilon^{\mu\nu\rho\sigma}\Gamma_\sigma\Gamma_5
\ee
the Lagrangian becomes
\be
\mathcal L_{gravitino}=\frac{i}{2}\bar\Psi_k\Gamma^{klm}\tilde{\mathcal D}_l\Psi_m+\frac{1}{2}m_\psi\bar\Psi_a\Gamma^{ab}\Psi_b
\ee
Writing in terms of $\gamma$-matrix ($\gamma_\mu=i\Gamma_\mu$), we get
\be
\Gamma^{\mu\nu\rho}=i\gamma^{\mu\nu\rho}
\ee
and
\be
\mathcal L_{gravitino}=-\frac{1}{2}\bar\Psi_\mu\gamma^{\mu\nu\rho}\tilde{\mathcal D}_\nu\Psi_\rho-\frac{1}{2}m_\psi\bar\Psi_\mu\gamma^{\mu\nu}\Psi_\nu
\ee
The susy transformation becomes
\be
\delta\Psi_\mu=2D_\mu\hat\epsilon+m_\psi\gamma_\mu\hat\epsilon
\ee
To calculate the gravitino partition function we will follow appendix A of \cite{Hoover:2005uf}.
We consider the following field redefinition. The motivation for this will be clear later.
\be
\Psi_\mu=\eta_\mu+A\gamma_\mu\eta,\quad \eta=\gamma^\mu\eta_\mu,\quad \bar\eta=\bar\eta_\mu\gamma^\mu
\ee
A is a real constant to be determined later. It is easy to see that the above field redefinitions have trivial Jacobian.
Now
\be
\gamma^\mu\Psi_\mu=(1+4A)\eta,\quad
\bar\Psi_\mu=\bar\eta_\mu+A\bar\eta\gamma_\mu\implies \bar\Psi_\mu\gamma^\mu=(1+4A)\bar\eta
\ee
We find that
\ben
\bar\Psi_\mu\gamma^{\mu\nu\rho}\tilde{\mathcal D}_\nu\Psi_\rho&=&\bar\eta\slashed D\eta\l[(1+4A)^2-2A(1+4A)-2A^2-2A\r]-(1+2A)\bar\eta \tilde{\mathcal D}^\mu\eta_\mu\nn\\&&-(1+2A)\bar\eta_\mu \tilde{\mathcal D}^\mu\eta-g^{\mu\nu}\bar\eta_\mu\gamma^\kappa\tilde{\mathcal D}_\kappa \eta_\nu
\een
Therefore choosing $A=-\frac{1}{2}$, the cross terms disappear and we get
\be
\bar\Psi_\mu\Gamma^{\mu\nu\rho}\tilde{\mathcal D}_\nu\Psi_\rho=\frac{1}{2}\bar\eta\slashed D\eta+g^{\mu\nu}\bar\eta_\mu\gamma^\kappa \tilde{\mathcal D}_\kappa \eta_\nu
\ee
Also
\be
\bar\Psi_\mu\Gamma^{\mu\nu}\Psi_\nu=\bar\eta\eta-g^{\mu\nu}\bar\eta_\mu\eta_\nu
\ee
The gravitino Lagrangian becomes
\be
\mathcal L_{gravitino}=-\frac{1}{4}\bar\eta\l(\slashed D+2m_\psi\r)\eta-\frac{1}{2}g^{\mu\nu}\bar\eta_\mu\l(\gamma^\kappa\tilde{\mathcal D}_\kappa-m_\psi\r)\eta_\nu
\ee
We also need to add gauge fixing condition. We put gauge condition $\eta=0$ and gauge fixing Lagrangian
\be
\mathcal L_{g.f.}=\frac{1}{4}\bar\eta\l(\slashed D+2m_\psi\r)\eta
\ee
This choice of gauge fixing Lagrangian introduces a determinant $det^{-1}\l(\slashed D+2m_\psi\r)$.
The Lagrangian becomes
\be
\mathcal L_{gravitino}+\mathcal L_{g.f.}=-\frac{1}{2}g^{\mu\nu}\bar\eta_\mu\l(\gamma^\kappa\tilde{\mathcal D}_\kappa-m_\psi\r)\eta_\nu
\ee

The corresponding supersymmtery transformation is
\be
\delta\eta=-\gamma^\mu\delta\Psi_\mu=-2\l(\slashed D+2m_\psi\r)\epsilon
\ee
which will give Fadeev Popov determinant $\sim det^{-2}\l(\slashed D+2m_\psi\r)$.\\
Therefore the complete partition function of Dirac gravitino is
\be
\mathcal Z_{Diracgravitino}\sim\frac{det\l(\gamma^\kappa\tilde{\mathcal D}_\kappa-m_\psi\r)|_{\eta_m}}{det^{3}\l(\slashed D+2m_\psi\r)|_\eta}
\ee
We have already calculated the coefficient of log correction from Dirac fermion. We here calculate the contribution from numerator. Now 
\be
\l(\gamma^\kappa\tilde{\mathcal D}_\kappa-m_\psi\r)\l(-\gamma^\mu\tilde{\mathcal D}_\mu-m_\psi\r)\eta_\rho=-\Box\eta_\rho+\frac{1}{4}R\eta_\rho-\frac{1}{2}\gamma^\mu\gamma^\nu R_{\mu\nu\rho\sigma}\eta^\sigma+m_\psi^2\eta_\rho
\ee
Thus we get the  deWitt coefficient,
\ben
\boldsymbol{a}_{4(gravitino)}&=&\frac{1}{360}\Big[-960R\l(\frac{1}{4}R+m_\psi^2\r)+2880\l(\frac{1}{4}R+m_\psi^2\r)^2\nn\\&&+212R_{\mu\nu\rho\sigma}R^{\mu\nu\rho\sigma}+80R^2-32R_{\mu\nu}R^{\mu\nu}\Big]
\een
We also need to include ghost contribution. The contribution from ghost is thrice the contribution of a massive Dirac fermion. The deWitt coefficient including the mass term for the ghost $\boldsymbol{a}_{4(ghost)}$ is given by
\ben
\boldsymbol{a}_{4(ghost)}=&=&\frac{12}{360}\Big[-60R\l(\frac{1}{4}R+4m_\psi^2\r)+180\l(\frac{1}{4}R+4m_\psi^2\r)^2\nn\\&&-\frac{7}{4}R_{\mu\nu\rho\sigma}R^{\mu\nu\rho\sigma}+5R^2-2R_{\mu\nu}R^{\mu\nu}\Big]
\een
Thus deWitt coefficient including the mass term for the physical gravitino is given by
\ben
\boldsymbol{a}_{4}(total)&=&\boldsymbol{a}_{4(gravitino)}-\boldsymbol{a}_{4(ghost)}\nn\\&&=\frac{1}{360}\Big[5R^2-960Rm_\psi^2-31680m_\psi^4+233R_{\mu\nu\rho\sigma}R^{\mu\nu\rho\sigma}\nn\\&&-8R_{\mu\nu}R^{\mu\nu}\Big]
\een
From the above expression for $\boldsymbol{a}_{4}(total)$, we can extract the coefficients in (\ref{GilkCoef1}) for the physical gravitino,
\ben
&&\boldsymbol{a}_{0}=-166,\quad \boldsymbol{a}_{2}=\frac{16}{6}R\nn\\&&\boldsymbol{a}_{4}=\frac{1}{360}\l[\frac{251}{6}R^2+233C_{\mu\nu\rho\sigma}C^{\mu\nu\rho\sigma}+458\l(R_{\mu\nu}R^{\mu\nu}-\frac{1}{4}R^2\r)\r]
\een
In the above we have calculated for Dirac gravitino, so to extract the contribution for Weyl/Majorana gravitino, we will divide the above results by half. 


\end{document}